\documentclass[10pt]{article}

\usepackage{amssymb}
\usepackage{amsmath}
\usepackage{graphics,graphicx}

\newcommand{\beq}{\begin{equation}}
\newcommand{\eeq}{\end{equation}}
\newcommand{\lb}{\label}
\newcommand{\beqar}{\begin{eqnarray}}
\newcommand{\eeqar}{\end{eqnarray}}
\newcommand{\bit}{\begin{itemize}}
\newcommand{\eit}{\end{itemize}}
\newcommand{\barr}{\begin{array}}
\newcommand{\earr}{\end{array}}
\def\ds{\displaystyle}

\newcommand{\modI}{\text{I}}
\newcommand{\modII}{\text{II}}
\newcommand{\modIII}{\text{III}}

\def\XXint#1#2#3{{\setbox0=\hbox{$#1{#2#3}{\int}$}
   \vcenter{\hbox{$#2#3$}}\kern-.5\wd0}}

\def\b0{\mbox{\boldmath $0$}}

\def\be{\mbox{\boldmath $e$}}

\def\bE{\mbox{\boldmath $E$}}

\def\bG{\mbox{\boldmath $G$}}

\def\bJ{\mbox{\boldmath $J$}}

\def\bR{\mbox{\boldmath $R$}}

\def\bU{\mbox{\boldmath $U$}}

\def\bY{\mbox{\boldmath $Y$}}

\def\f0{\ensuremath{\mathbb{O}}}

\newcommand{\mA}{\ensuremath{\mathcal{A}}}
\newcommand{\mB}{\ensuremath{\mathcal{B}}}

\newcommand{\mF}{\ensuremath{\mathcal{F}}}

\def\Im{\mathop{\mathrm{Im}}}
\def\Re{\mathop{\mathrm{Re}}}

\newcommand{\sign}{\mathop{\mathrm{sign}}}

\newcommand{\Reals}{\ensuremath{\mathbf{R}}}
\newcommand{\Complex}{\ensuremath{\mathbf{C}}}

\def\IJSS{{\it Int.\ J.\ Solids Struct.}\ }

\def\JAM{{\it ASME J.\ Appl.\ Mech.}\ }

\def\JMPS{{\it J.\ Mech.\ Phys.\ Solids}\ }

\begin{document}

\title{Evaluation of the Lazarus-Leblond constants in the asymptotic model of the interfacial wavy crack}

\author{A. Piccolroaz$^a$, G. Mishuris$^b$, A.B. Movchan$^c$
\\
\\
$^a$
{\it Dipartimento di Ingegneria Meccanica e Strutturale, Universit`a di Trento,}
\\ {\it Via Mesiano 77, I-38050 Trento, Italia}
\\
{\it
$^b$
Department of Mathematics, Rzeszow University of Technology,}
\\ {\it W. Pola 2, 35-959, Rzeszow, Poland,}
\\
{\it
$^c$
Department of Mathematical Sciences, University of Liverpool,}
\\ {\it Liverpool L69 3BX, U.K.}
}

\maketitle

\begin{abstract}
\noindent
The paper addresses the problem of a a semi-infinite plane crack along the
interface between two 3D isotropic half-spaces. Two methods of solution have
been considered in the past: Lazarus and Leblond (1998) applied the
``special'' method by Bueckner (1987) and found the expression of the
variation of the stress intensity factors for a wavy crack without solving the
complete elasticity problem; their solution is expressed in terms of the
physical variables,
and it involves five constants whose analytical representation was
unknown; on the other hand the ``general'' solution to the problem
has been recently addressed by Bercial-Velez {\em et al.}\ (2005),
using a Wiener-Hopf analysis and singular asymptotics near the
crack front.

The main goal of the present paper is to complete the solution to
the problem by providing the connection between the two methods.
This is done by constructing an integral representation for the
Lazarus-Leblond's weight functions and by deriving the closed form
representations of the Lazarus-Leblond's constants.
\end{abstract}

\vspace{5mm} \noindent{\it Keywords:} stress intensity factor;
interfacial crack; Betti formula; weight function.

\newpage

\tableofcontents

\newpage

\section{Introduction.}

The fundamentals of the theory of three-dimensional interfacial cracks and the 
corresponding integral equation formulations were introduced by Willis (1971a, 1971b).
A singularly perturbed problem of a general static loading of a semi-infinite plane
crack
along the
interface
between two isotropic half-spaces of different linear
elastic
materials is
considered here. The two
characteristic features of this problem are the coupling of all
three opening modes and the oscillatory behaviour of the solution
near the crack edge. For this problem two issues are of major
interest: the representation of the stress intensity factors, and
the expression of the variation of stress intensity factors
arising from an infinitesimal coplanar perturbation of the crack
front. Two methods of solution have been considered in the past:
Lazarus and Leblond (1998a, 1998b) applied the special method
by Bueckner (1987) and found the expression of the variation of
the stress intensity factors without solving the complete
elasticity problem; their perturbation formulae are simple and
elegant, expressed in terms of the physical variables, but
these formulae involve five constants whose analytical
representation was left unknown; on the other hand the
general
solution to the problem has been independently  addressed by
Antipov (1999) and Bercial-Velez {\em et al.}\ (2005).

In the present paper, we revisit the work by Bercial-Velez
{\em et  al.}\ (2005) and show that it is possible to
construct
the weight
functions in the form directly
related to the weight
functions of Lazarus and Leblond (1998a). This also yields the
required analytical expressions for the unknown constants in
asymptotic formulae of Lazarus and Leblond (1998a). Then, using an
asymptotic method and the integral reciprocal
identity, formerly introduced by Willis and Movchan (1995),
we obtain canonical integral representations for the stress intensity
factors.

The structure of the paper is as follows. Section 2
includes the governing equations and presents the main result. The
fundamental identity and the weight functions are described in
Section 3. Closed form representations for the Fourier
transforms of the weight functions are given in Section 4.
Local asymptotics and comparison with the Lazarus-Leblond's weight
functions are given in Section 5. Finally, Section
6 includes a brief outline of applications to the
wavy crack problem.

\section{Problem definition and main result.}
\begin{figure}[!htb]
\begin{center}
\vspace*{3mm}
\includegraphics[width=12cm]{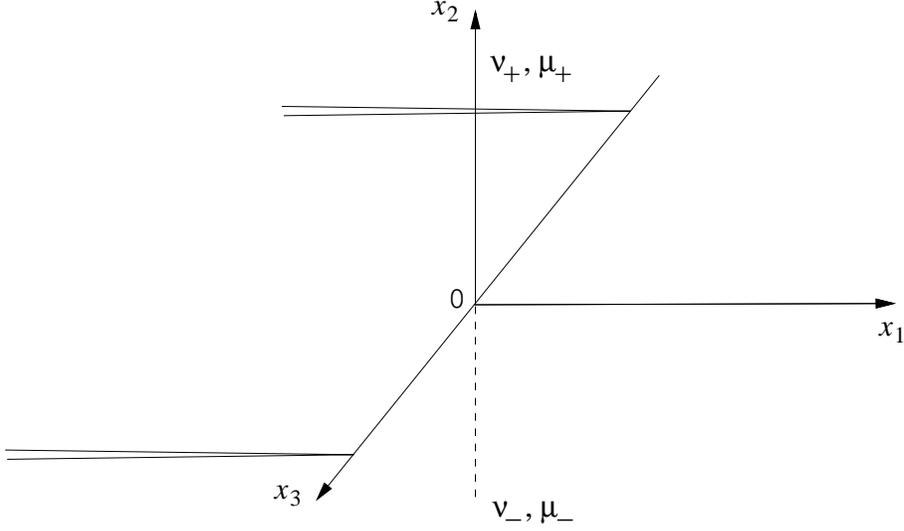}
\caption{\footnotesize Domain for the physical solution.}
\label{fig01}
\end{center}
\end{figure}

\lb{result}
We consider an infinite elastic body, consisting of
two different isotropic materials that occupy the upper half-space
(material 1)
$\Reals_3^+ = \{ (x_1, x_2, x_3) \in \Reals_3: x_2 > 0\}$ and the lower
half-space (material 2)
$\Reals_3^- = \{ (x_1,x_2, x_3) \in \Reals_3: x_2 < 0\}$ (see Fig.~\ref{fig01}).
The
Poisson's ratio and the shear modulus of materials 1 and 2 are
denoted by  $\nu_+$, $\mu_+$ and  $\nu_-$, $\mu_-$, respectively.

The crack lies on the half-plane
$\Reals_2^- = \{ (x_1,x_2,x_3) \in \Reals_3: x_1 < 0, x_2 =0 \}$, whereas the
ideal contact conditions
are valid
on the half-plane
$\Reals_2^+ = \{ (x_1,x_2,x_3) \in \Reals_3: x_1 > 0, x_2 = 0 \}$. This
implies that the traction and displacement components
are continuous across the interface ahead of the crack front
\beq
[\sigma_{12}] = [\sigma_{22}] = [\sigma_{32}] = 0, \qquad
[u_1] = [u_2] = [u_3] = 0, ~~\mbox{as} ~ x_1 > 0,
\eeq
where the square
brackets denote
the jump while crossing the interface,
\beq
[f](x_1,x_3) =
f(x_1,+0,x_3) - f(x_1,-0,x_3).
\eeq
The Stress Intensity Factors $K_j, ~ j=\modI, \modII, \modIII,$ are defined as
in Hutchinson {\em et al.}\ (1987)
(also see Lazarus and Leblond (1998)), so that the asymptotics of
tractions ahead of the crack edge and the asymptotics of the
displacement jump across the crack surfaces take the form
\beq
\left\{
\barr{l}
\ds \sigma_{22}(x_1,0,x_3) + i
\sigma_{12}(x_1,0,x_3) \sim
\frac{K(x_3)}{\sqrt{2\pi x_1}} x_1^{i \epsilon}, \quad x_1 \to 0^+, \\[5mm]
\ds
\sigma_{23}(x_1,0,x_3) \sim
\frac{K_{\modIII}(x_3)}{\sqrt{2\pi x_1}}, \quad x_1 \to 0^+,
\earr
\right.
\eeq
and
\beq
\left\{
\barr{l}
\ds
[u_2 + i u_1](x_1,x_3) \sim
\frac{(1-\nu_+)/\mu_+ + (1-\nu_-)/\mu_-}{(1/2+i\epsilon) \cosh(\pi\epsilon)}
K(x_3)
\sqrt{\frac{-x_1}{2\pi}} (-x_1)^{i\epsilon}, \quad x_1 \to 0^-, \\[5mm]
\ds
[u_3](x_1,x_3) \sim
2\left(\frac{1}{\mu_+}+\frac{1}{\mu_-}\right)
K_{\modIII}(x_3) \sqrt{\frac{-x_1}{2\pi}}, \quad x_1 \to 0^-,
\earr
\right.
\eeq
respectively, where $K(x_3) = K_{\modI}(x_3) + iK_{\modII}(x_3)$ is the
complex stress intensity factor and
\beq
\lb{bimat}
\epsilon = \frac{1}{2\pi}
\log\frac{\mu_+ + (3-4\nu_+)\mu_-}{\mu_- + (3-4\nu_-)\mu_+}
\eeq
is the bimaterial constant.

The crack faces are loaded by
surface tractions with
components $-p_i(x_1,x_3)$ on the upper face ($x_2 = 0^+$) and $p_i(x_1,x_3)$
on the lower face ($x_2 = 0^-$).

The body forces are assumed to be zero, and the displacement
components satisfy
the Lam\'{e} system
\beq
\lb{lame}
u_{k,ik} + (1-2\nu_{\pm}) u_{i,jj} = 0, \quad \text{in } \Reals_3^{\pm},
\eeq
with boundary
conditions $\sigma_{i2}(x_1,x_2,x_3) = p_i(x_1,x_3)$ as $x_2 \to
0^{\pm}$, on the crack surfaces ($x_1 < 0$), and ideal contact
conditions $[u_i](x_1,x_3) = [\sigma_{i2}](x_1,x_3) = 0$ on the
interface ($x_1 > 0$).

Lazarus and Leblond (1998) derived expressions for the variation
of SIFs generated by a small coplanar perturbation of the crack
front, $x_1 = \delta \phi(x_3)$.
In terms of the  Fourier transforms and assuming that $K$ and $K_{\modIII}$
are initially uniform, Lazarus-Leblond's formulae
are written as follows, \footnote{These expressions were kindly
supplied to us by J.B. Leblond.}
\beqar
\lefteqn{
\widetilde{\Delta K}(\lambda) = \widetilde{\frac{dK}{da} \delta\phi} -
\frac{1+2i\epsilon}{8\cosh(\pi\epsilon)} \left\{ \gamma_+
\frac{\sinh(\pi\epsilon)}{\epsilon} \Gamma(1-2i\epsilon) K^*
\frac{|\lambda|^{1+2i\epsilon}}{1+2i\epsilon} + \pi\gamma_- K
|\lambda|
\right. } \nonumber \\[5mm]
&& \hspace{30mm}
\left.
- \frac{4\gamma_{\modIII}}{1-\nu}
\frac{\cosh(\pi\epsilon/2)}{\epsilon}
\Gamma(1-i\epsilon)K_{\modIII} \sign(\lambda)
\frac{|\lambda|^{1+i\epsilon}}{1+i\epsilon} \right\}
\widetilde{\delta\phi},
\eeqar
\beq
\widetilde{\Delta K}_{\modIII}(\lambda) =
\widetilde{\frac{dK_{\modIII}}{da} \delta\phi} -\left\{
\frac{\pi\gamma}{4} K_{\modIII} |\lambda| -
i\frac{1-\nu}{2}\frac{\cosh(\pi\epsilon/2)}{\epsilon}
\sign(\lambda) \Im \left[ \gamma_z \Gamma(1-i\epsilon) K^*
\frac{|\lambda|^{1+i\epsilon}}{1+i\epsilon} \right]
\right\}\widetilde{\delta\phi}, \nonumber
\eeq
where the tilde denotes Fourier transform with respect to $x_3$, the
star denotes complex conjugation, $dK/da$ and $dK_{\modIII}/da$ represent the
derivative of $K$ and $K_{\modIII}$, respectively, for
a uniform advance of the crack front and $\Gamma$ is the gamma function.

Here $\gamma_+$, $\gamma_-$,
$\gamma_{\modIII}$, $\gamma_{z}$ are complex constants, and
$\gamma$ is a real constant, and the analytical representations
for these constants were absent in the literature. We note that
the constants $\gamma_+$, $\gamma_-$, $\gamma_{\modIII}$,
$\gamma_{z}$, $\gamma$ are taken from the asymptotics of the
weight functions used by Lazarus and Leblond (1998).

The main aim of the present paper is to fill-in this gap and to
obtain the representation for the Lazarus-Leblond's constants. This
was possible due to the analytical results of Antipov (1999) and
Bercial-Velez {\em et al.}\ (2005). The required formulae have the form
\beq
\lb{gm1}
\gamma = \frac{2}{\pi}
\frac{3(b+e)-\sqrt{b^2-d^2}}{\sqrt{b^2-d^2}+b+e}, \nonumber
\eeq
\beq
\gamma_{\modIII} =
-\frac{8\sqrt{\pi}\epsilon(1+i\epsilon)\sqrt{b^2-d^2}\sqrt[4]{b^2-d^2}}
{2^{i\epsilon}(1+2i\epsilon)\Gamma(1/2+i\epsilon)\Gamma(1-i\epsilon)
(\sqrt{b^2-d^2}+b+e)(\sqrt{b+d}+\sqrt{b-d})}, \nonumber
\eeq
\beq
\gamma_z = - \gamma_{\modIII} (1+2i\epsilon)
\frac{b+e}{\sqrt{b^2-d^2}},
\eeq
\beq
\gamma_- = \frac{8b}{\pi (1+2i\epsilon)(\sqrt{b^2-d^2}+b+e)}, \nonumber
\eeq
\beq
\lb{gm5}
\gamma_+ = -\frac{4\epsilon
b\Gamma(1/2-i\epsilon)(\sqrt{b^2-d^2}-b-e)}
{4^{i\epsilon}d\Gamma(1/2+i\epsilon)\Gamma(1-2i\epsilon)(\sqrt{b^2-d^2}+b+e)},
\nonumber
\eeq
where
\beq
\lb{const}
b = \frac{1-\nu_+}{\mu_+} + \frac{1-\nu_-}{\mu_-}, \quad
d = \frac{1-2\nu_+}{2\mu_+} - \frac{1-2\nu_-}{2\mu_-}, \quad
e = \frac{\nu_+}{\mu_+} + \frac{\nu_-}{\mu_-}.
\eeq
Note that, using the parameters in (\ref{const}), the bimaterial constant
$\epsilon$ given by (\ref{bimat}) can be written as
\beq
\epsilon = \frac{1}{2\pi} \log\frac{b+d}{b-d}.
\eeq

\section{The fundamental identity.}
\lb{fund_id}
We will adapt the method introduced for the
homogeneous elastic space by Willis and Movchan (1995). This
method involves the use of the reciprocal theorem (Betti formula)
in order to relate the physical solution to the weight function,
which is a special singular solution to the homogeneous problem.

In the absence of body forces, the Betti formula takes the form
\beq
\int_{\partial\Omega} \{ \sigma_{ij}^{(1)} n_j u_i^{(2)} -
\sigma_{ij}^{(2)} n_j u_i^{(1)} \} ds = 0,
\eeq
where
$\partial\Omega$ is any surface enclosing a region $\Omega$ within
which both $u_i^{(1)}$ and $u_i^{(2)}$ satisfy the equations of
equilibrium (\ref{lame}), with corresponding stress states
$\sigma_{ij}^{(1)}$ and $\sigma_{ij}^{(2)}$, and $n_i$ denotes the
outward normal to $\partial\Omega$.

Applying the Betti formula to a hemispherical domain in the upper
half-space $\Reals_3^+$, whose plane boundary is $x_2 = 0^+$ and
whose radius $R$ will be allowed to tend to infinity, we obtain,
in the limit $R \to \infty$,
\beq
\lb{betti}
\int\limits_{(x_2=0^+)} \{ \sigma_{i2}^{(1)}(x_1,0^+,x_3)
u_i^{(2)}(x_1,0^+,x_3) - \sigma_{i2}^{(2)}(x_1,0^+,x_3)
u_i^{(1)}(x_1,0^+,x_3) \} dx_1 dx_3 = 0,
\eeq
provided that the
fields $u_i^{(1)}$ and $u_i^{(2)}$ decay suitably fast at
infinity. We can also assume that $u_i^{(2)}$ represents a
non-trivial solution of the homogeneous problem, whereas $u_i^{(1)}$ stands
for the physical field associated with the crack
loaded at its surface.

\begin{figure}[!htb]
\begin{center}
\vspace*{3mm}
\includegraphics[width=9cm]{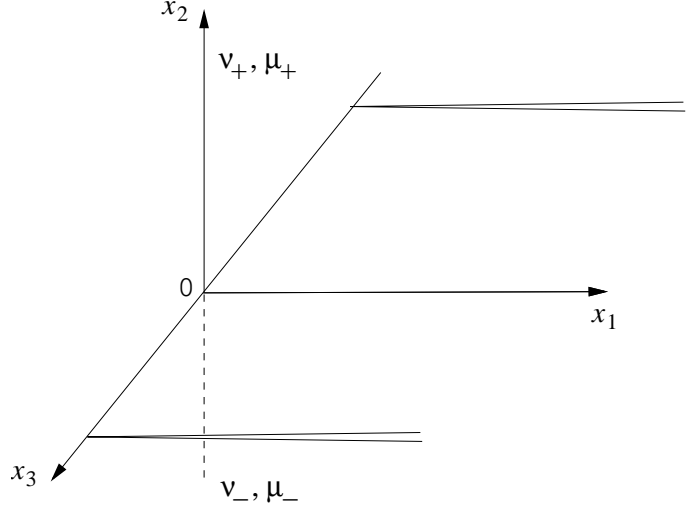}
\caption{\footnotesize Domain for the weight functions.}
\label{fig02}
\end{center}
\end{figure}

Similar to Bercial-Velez {\em et al.}\ (2005), we now define a new
vector function $\{U_i\}_{i=1}^3$ in the following way
\beq
\lb{trans}
\barr{l}
U_1(-x_1,x_2,-x_3) = - u_1^{(2)}(x_1,x_2,x_3), \\[5mm]
U_2(-x_1,x_2,-x_3) =   u_2^{(2)}(x_1,x_2,x_3), \\[5mm]
U_3(-x_1,x_2,-x_3) = - u_3^{(2)}(x_1,x_2,x_3),
\earr
\eeq
which
corresponds to introducing a change of coordinates in the solution
$u_i^{(2)}$, namely a rotation\footnote{It is noted that this
transformation of coordinates differs from the transformation of
reflection used by Willis and Movchan (1995) for the case of a
crack in a homogeneous elastic space.} about the $x_2$-axis
through an angle $\pi$.
It is straightforward to verify that
the function $U_i$
satisfies the equations of equilibrium (\ref{lame}), but in a
different domain (see Fig.~\ref{fig02}). In the sequel, the vector function
$\{U_i\}_{i=1}^3$ will play the role of the weight function, whereas the
vector function $\{u_i^{(1)}\}_{i=1}^3$ will be identified with the physical
solution (and we will drop the superscript $(1)$, no longer needed
in the following notations). The notation
$\Sigma_{hk}(X_1,X_2,X_3)$ will be used for components of stress
corresponding to the displacement field $U_i (X_1,X_2,X_3)$.
Equivalently (see (\ref{trans})), the components of displacement
and stress are related by the formulae
\beq
u_i^{(2)}(x_1,x_2,x_3)
= R_{ih} U_h(-x_1,x_2,-x_3),
\eeq
and
\beq
\sigma_{ij}^{(2)}(x_1,x_2,x_3) = R_{ih} \Sigma_{hk}(-x_1,x_2,-x_3)
R_{kj},
\eeq
where
\beq
\bR =
\left(
\barr{ccc}
-1 & 0 &  0 \\
 0 & 1 &  0 \\
 0 & 0 & -1
\earr
\right).
\eeq

Replacing  $u_i^{(2)}(x_1, x_2, x_3)$ with $u_i^{(2)}(x_1-x_1',
x_2, x_3-x_3')$, which corresponds to a shift within the plane
$(x_1, x_3),$ we obtain
\beqar
\label{eq20}
\lefteqn{
\int\limits_{(x_2=0^+)}
\{ \sigma_{i2}(x_1,0^+,x_3) R_{ih} U_h(x_1'-x_1,0^+,x_3'-x_3) }
\nonumber \\[0mm]
&& \hspace{30mm}
- R_{ih} \Sigma_{h2}(x_1'-x_1,0^+,x_3'-x_3)
u_i(x_1,0^+,x_3) \} dx_1 dx_3 = 0.
\eeqar

A similar equation can be derived by applying the Betti formula to a
hemispherical domain in the lower half-space $\Reals_3^-$,
\beqar
\label{eq21}
\lefteqn{
\int\limits_{(x_2=0^-)} \{ \sigma_{i2}(x_1,0^-,x_3)
R_{ih} U_h(x_1'-x_1,0^-,x_3'-x_3) }
\nonumber \\[0mm]
&& \hspace{30mm}
- R_{ih} \Sigma_{h2}(x_1'-x_1,0^-,x_3'-x_3)
u_i(x_1,0^-,x_3) \} dx_1 dx_3 = 0,
\eeqar
and hence by
subtraction of (\ref{eq21}) from (\ref{eq20}), we obtain
\beqar
\lefteqn{
\int\limits_{(x_2=0)} \{ \sigma_{i2}(x_1,0,x_3) R_{ih}
[U_h](x_1'-x_1,x_3'-x_3) }
\nonumber \\[0mm]
&& \hspace{30mm}
- R_{ih} \Sigma_{h2}(x_1'-x_1,0,x_3'-x_3)
[u_i](x_1,x_3) \} dx_1 dx_3 = 0,
\eeqar
which is a convolution
integral and the equivalent representation is
\beq
\{ \sigma_{i2} * R_{ih} [U_h] - R_{ih} \Sigma_{h2} * [u_i] \}(x_1',x_3') = 0.
\eeq

Upon writing
\beq
\sigma_{i2}(x_1,0,x_3) = p_i(x_1,x_3) + \sigma_{i2}^{(+)}(x_1,x_3),
\eeq
where $p_i(x_1,x_3) = \sigma_{i2}(x_1,0,x_3) H(-x_1)$ is the loading, and
$\sigma_{i2}^{(+)}(x_1,x_3) = \sigma_{i2}(x_1,0,x_3) H(x_1)$ is the stress
field ahead of the crack edge, with $H(x_1)$ being the Heaviside function, we
get the Betti identity in the following form
\beq
\lb{ident}
\{ \sigma_{i2}^{(+)} * R_{ih} [U_h] - R_{ih} \Sigma_{h2} * [u_i] \}(x_1',x_3')
= - \{ p_i * R_{ih} [U_h] \}(x_1',x_3').
\eeq
Equation (\ref{ident})
will be used for evaluation of  the stress intensity factors. A
field with components $U_i(x_1,x_2,x_3)$
is the weight function defined as follows
\bit
\item[(a)]
it satisfies the equation of equilibrium (\ref{lame});
\item[(b)]
$[U_i] = 0$ when $x_1 < 0$;
\item[(c)]
the associated
$\Sigma_{i2}$ is continuous and $\Sigma_{i2} = 0$ when $x_2 = 0$ and $x_1 > 0$
(homogeneous boundary conditions);
\item[(d)]
$[U_i] \sim k_i(x_1) x_1^{-1/2} \delta(x_3)$ as $x_1 \to 0^+$, where
$k_i(x_1)$ is a bounded function, and    $\delta(x_3)$ denotes the Dirac
delta function.\footnote{
The bounded functions $k_i(x_1)$ have the form
$
k_i(x_1) = k_i^{(1)} x_1^{i\epsilon}
+ k_i^{(2)} x_1^{-i\epsilon} + k_i^{(3)},
$
where $k_i^{(1)},k_i^{(2)},k_i^{(3)}$ are constants.}
\item[(e)]
$U_i$ is a linear combination of homogeneous functions of degree
$-3/2$, $-3/2 + i\epsilon$ and $-3/2 - i\epsilon$.
\eit

\section{The weight functions.}
\lb{WF}
Let us introduce the Fourier transforms of the displacement jump
and of the traction components for the weight functions
\beq
\barr{l}
\ds [\overline{U}_i]^+(\beta,\gamma) =
\int_{-\infty}^{\infty} \int_{0}^{\infty} [U_i](x_1,x_3)
e^{i\beta x_1 + i\lambda x_3} dx_1 dx_3, \\[5mm]
\ds \overline{\Sigma}^-_{i2}(\beta,\gamma) =
\int_{-\infty}^{\infty} \int_{-\infty}^{0} \Sigma_{i2}(x_1,x_3)
e^{i\beta x_1 + i\lambda x_3} dx_1 dx_3.
\earr
\eeq
The weight
functions are defined in such a way that $[\overline{U}_i]^+$ is
analytic in the upper half-plane
\beq
\Complex_{\beta}^+ =
\{-\infty<\Re(\beta)<\infty, \Im(\beta)>0\},
\eeq
whereas
$\overline{\Sigma}_{i2}^-$ is analytic in the lower half-plane
\beq
\Complex_{\beta}^- = \{-\infty<\Re(\beta)<\infty,
\Im(\beta)<0\}.
\eeq

\subsection{The Wiener-Hopf problem.}
The relationship between the Fourier transforms of displacement jump and
traction components has been derived by Willis (1971a, 1971b) and takes the
form
\beq
\lb{wh}
[\overline{\bU}]^+(\beta,\lambda) =
\frac{1}{\rho}
\bG(\beta,\lambda) \overline{\mbox{\boldmath$\Sigma$}}^-(\beta,\lambda),
\eeq
where $\lambda \in \Reals$ and $[\overline{\bU}]^+(\beta,\lambda)$,
$\overline{\mbox{\boldmath$\Sigma$}}^-(\beta,\lambda)$ are the limit values of
the functions as $\Im(\beta) \to 0$,
\beq
\barr{l}
\rho = \sqrt{\beta^2 + \lambda^2}, \quad
\bG = - \frac{1}{\rho^2}
\left(
\barr{ccc}
-id \beta \rho   & b\rho^2 + e\lambda^2 & -e \beta \lambda     \\[5mm]
b \rho^2         & id \beta \rho        & id \lambda \rho      \\[5mm]
-id \lambda \rho & -e \beta \lambda     & b \rho^2 + e \beta^2
\earr
\right),
\\[13mm]
\ds
b = \frac{1-\nu_+}{\mu_+} + \frac{1-\nu_-}{\mu_-}, \quad
d = \frac{1-2\nu_+}{2\mu_+} - \frac{1-2\nu_-}{2\mu_-}, \quad
e = \frac{\nu_+}{\mu_+} + \frac{\nu_-}{\mu_-}.
\earr
\eeq
Without loss of generality we assume $d > 0$.

Note that the order of the components of the traction vector
$\overline{\mbox{\boldmath$\Sigma$}}^-$ in equation (\ref{wh}) is different from
the standard one, namely
\beq
\overline{\mbox{\boldmath$\Sigma$}}^- =
\left(
\overline{\Sigma}^-_{22},
\overline{\Sigma}^-_{12},
\overline{\Sigma}^-_{32}
\right)^T.
\eeq

Following Antipov (1999), the matrix $\bG$ can be written as
$\bG = \bJ_1 \bG_0 \bJ_2$, where
\beq
\barr{l}
\bJ_1 = \bJ_1^{-T} =
\left(
\barr{ccc}
0 & 0 & 1 \\
1 & 0 & 0 \\
0 & 1 & 0
\earr
\right), \quad
\bJ_2 = \bJ_2^{-1} =
\left(
\barr{ccc}
1 & 0 & 0 \\
0 & 0 & 1 \\
0 & 1 & 0
\earr
\right),
\\[8mm]
\ds
\bG_0 = - \frac{1}{\rho^2}
\left(
\barr{ccc}
b \rho^2         & id \lambda \rho      & id \beta \rho          \\[5mm]
-id \lambda \rho & b \rho^2 + e \beta^2 & -e \beta \lambda       \\[5mm]
-id \beta \rho   & -e \beta \lambda     & b \rho^2 + e \lambda^2
\earr
\right).
\earr
\eeq

Here we use the following normalization with respect to mechanical parameters
\beq
\mbox{\boldmath$\phi$}^+(\beta,\lambda) = \bJ_1^{-1}
[\overline{\bU}]^+(\beta,\lambda), \quad
\mbox{\boldmath$\phi$}^-(\beta,\lambda) = b \bJ_2
\overline{\mbox{\boldmath$\Sigma$}}^-(\beta,\lambda),
\eeq
so that
the problem can now be written as
\beq
\mbox{\boldmath$\phi$}^+(\beta,\lambda) = \frac{1}{\rho}
\bG_1(\beta,\lambda)
\mbox{\boldmath$\phi$}^-(\beta,\lambda), \label{eq37}
\eeq
where
\beq
\bG_1 = - \frac{1}{\rho^2} \left(
\barr{ccc}
\rho^2           & i d_* \lambda \rho & id_* \beta \rho \\[5mm]
-id_*\lambda\rho & \rho^2+e_*\beta^2  & -e_*\beta\lambda \\[5mm]
-id_*\beta\rho   & -e_*\beta\lambda   & \rho^2 + e_*\lambda^2
\earr
\right),
\eeq
and the dimensionless parameters $d_*$ and $e_*$ are defined as follow:
\beq
d_* = \frac{d}{b} < 1,
\quad e_* = \frac{e}{b}.
\eeq
Using a new variable $\xi =
\beta / |\lambda|$, we can write equation (\ref{eq37}) in the form
\beq
\lb{whprob}
\mbox{\boldmath$\phi$}_*^+(\xi) =
\frac{1}{\rho_*} \bG_*(\xi,\sign(\lambda))
\mbox{\boldmath$\phi$}_*^-(\xi),
\eeq
where
\beq
\barr{l}
\ds
\rho_* =
\sqrt{\xi^2 + 1}, \quad
\bG_* = - \frac{1}{\rho_*^2} \left(
\barr{ccc}
\rho_*^2 & i d_* \sign(\lambda) \rho_* & id_* \xi \rho_* \\[5mm]
-id_*\sign(\lambda)\rho_* & \rho_*^2+e_*\xi^2
& -e_*\xi\sign(\lambda) \\[5mm]
-id_*\xi\rho_*  &-e_*\xi\sign(\lambda) &\rho_*^2 +
e_*
\earr
\right),
\\[12mm]
\ds
\mbox{\boldmath$\phi$}_*^+(\xi)
= \mbox{\boldmath$\phi$}^+(\xi |\lambda|,\lambda), \quad
\mbox{\boldmath$\phi$}_*^-(\xi) = \frac{1}{|\lambda|}
\mbox{\boldmath$\phi$}^-(\xi |\lambda|,\lambda).
\earr
\eeq
Note that
$\mbox{\boldmath$\phi$}_*^+$ is analytic in $\Complex_{\xi}^+$,
whereas $\mbox{\boldmath$\phi$}_*^-$ is analytic in
$\Complex_{\xi}^-$. The solution of the Wiener-Hopf equation (\ref{whprob}) is
outlined in the Appendix.

\subsection{Fourier transforms of the weight functions.}
The Fourier transforms of three linearly independent weight vector
functions and the corresponding traction components are:
\bit
\item[(1)] the first weight function
\beq
\lb{first}
\barr{l}
\ds \phi_{*3}^{+1} =
[\overline{U}_{*1}^1]^+ = \sign(\lambda) \left\{ \frac{1}{\xi-i} -
2\xi F_1(\xi)
\right\} (\xi+i)^{-1/2} , \\[5mm]
\ds
\phi_{*1}^{+1} = [\overline{U}_{*2}^1]^+ =
2 \sign(\lambda) F_2(\xi) (\xi+i)^{-1/2} , \\[5mm]
\ds \phi_{*2}^{+1} = [\overline{U}_{*3}^1]^+ = -\left\{
\frac{\xi}{\xi-i} + 2 F_1(\xi) \right\} (\xi+i)^{-1/2},
\earr
\eeq
\beq
\barr{l}
\ds \phi_{*3}^{-1} = \frac{b}{|\lambda|}
\overline{\Sigma}_{*12}^{-1} = -\sign(\lambda) \left\{
\frac{1}{(1+e_*)(\xi-i)} + 2id_*\xi P_1(\xi)
\right\} (\xi-i)^{1/2}  ,\\[5mm]
\ds
\phi_{*1}^{-1} = \frac{b}{|\lambda|} \overline{\Sigma}_{*22}^{-1} =
-2id_* \sign(\lambda) P_2(\xi) (\xi-i)^{1/2} , \\[5mm]
\ds \phi_{*2}^{-1} = \frac{b}{|\lambda|} \overline{\Sigma}_{*32}^{-1}
= \left\{ \frac{\xi}{(1+e_*)(\xi-i)} - 2id_* P_1(\xi)
\right\} (\xi-i)^{1/2},
\earr
\eeq

\item[(2)] the second weight function
\beq
\barr{l}
\ds
\phi_{*3}^{+2} = [\overline{U}_{*1}^2]^+ =
\left\{
\frac{1+e_*}{2d_0^2d_*(i-a)} + i\xi F_2(\xi)
\right\} (\xi+i)^{-3/2}, \\[5mm]
\ds
\phi_{*1}^{+2} = [\overline{U}_{*2}^2]^+ =
i\rho_*^2(\xi) F_1(\xi) (\xi+i)^{-3/2}, \\[5mm]
\ds \phi_{*2}^{+2} = [\overline{U}_{*3}^2]^+ = -\sign(\lambda)
\left\{ \frac{1+e_*}{2d_0^2d_*(i-a)}\xi - iF_2(\xi)
\right\} (\xi+i)^{-3/2},
\earr
\eeq
\beq
\barr{l}
\ds
\phi_{*3}^{-2} = \frac{b}{|\lambda|} \overline{\Sigma}_{*12}^{-2} =
\left\{ -\frac{1}{2d_0^2d_*(i-a)} + d_*\xi P_2(\xi)
\right\} \frac{(\xi-i)^{1/2}}{\xi+i}, \\[5mm]
\ds
\phi_{*1}^{-2} = \frac{b}{|\lambda|} \overline{\Sigma}_{*22}^{-2} =
-d_*\rho_*^2(\xi) P_1(\xi) \frac{(\xi-i)^{1/2}}{\xi+i} ,\\[5mm]
\ds \phi_{*2}^{-2} = \frac{b}{|\lambda|} \overline{\Sigma}_{*32}^{-2}
= \sign(\lambda) \left\{ \frac{\xi}{2d_0^2d_*(i-a)} +
d_*P_2(\xi) \right\} \frac{(\xi-i)^{1/2}}{\xi+i},
\earr
\eeq

\item[(3)] the third weight function
\beq
\lb{third}
\barr{l}
\ds \phi_{*3}^{+3} =
[\overline{U}_{*1}^3]^+ = \frac{\xi}{d_*(\xi-a)} \left\{
F_2(\xi) -\frac{i(i-\xi)}{d_*(i-a)} F_1(\xi)
\right\} (\xi+i)^{-1/2}, \\[5mm]
\ds
\phi_{*1}^{+3} = [\overline{U}_{*2}^3]^+ = \frac{1}{d_*(\xi-a)}
\left\{
\rho_*^2(\xi)F_1(\xi) +\frac{i(i-\xi)}{d_*(i-a)} F_2(\xi)
\right\} (\xi+i)^{-1/2}, \\[5mm]
\ds \phi_{*2}^{+3} = [\overline{U}_{*3}^3]^+ =
\frac{\sign(\lambda)}{d_*(\xi-a)} \left\{ F_2(\xi)
-\frac{i(i-\xi)}{d_*(i-a)} F_1(\xi) \right\} (\xi+i)^{-1/2},
\earr
\eeq
\beq
\barr{l}
\ds \phi_{*3}^{-3} = \frac{b}{|\lambda|}
\overline{\Sigma}_{*12}^{-3} = \left\{ -\frac{i\xi P_2(\xi)}{\xi-a}
+\frac{\xi P_1(\xi)}{d_*(\xi-a)} -\frac{\xi
P_1(\xi)}{d_*(i-a)}
\right\} (\xi-i)^{1/2}, \\[5mm]
\ds
\phi_{*1}^{-3} = \frac{b}{|\lambda|} \overline{\Sigma}_{*22}^{-3} =
\left\{
\frac{i\rho_*^2(\xi)P_1(\xi)}{\xi-a} +\frac{P_2(\xi)}{d_*(\xi-a)}
-\frac{P_2(\xi)}{d_*(i-a)}
\right\} (\xi-i)^{1/2}, \\[5mm]
\ds \phi_{*2}^{-3} = \frac{b}{|\lambda|} \overline{\Sigma}_{*32}^{-3}
= \sign(\lambda) \left\{ -\frac{i P_2(\xi)}{\xi-a}
+\frac{P_1(\xi)}{d_*(\xi-a)}
-\frac{P_1(\xi)}{d_*(i-a)} \right\} (\xi-i)^{1/2}.
\earr
\eeq
\eit
Here, the functions $F_j, P_j, ~ j=1,2,$ are defined by
\beq
\barr{l}
\ds
F_1(\xi) = -i\frac{M_-}{d_*}
\frac{\Lambda_*^+(\xi)\Sigma_{*1}^+(\xi)}{\rho_*^2(\xi)}, \quad
F_2(\xi) = -i\frac{M_-}{d_*}
\Lambda_*^+(\xi)\Sigma_{*2}^+(\xi),
\\[5mm]
\ds
P_1(\xi) = M_-
\frac{\Lambda_*^-(\xi)\sin B_{*}^-(\xi)}{\rho_*(\xi)}, \quad
P_2(\xi) = M_- \Lambda_*^-(\xi)\cos B_{*}^-(\xi).
\earr
\eeq

The functions $B_*^-, \Sigma_{* j}^+, \Lambda_*^\pm$ and the
constants $a, M_-, d_0$ are defined in the Appendix, equations (\ref{Bpm})$_2$,
(\ref{sigmap}), and (\ref{const2}).

\section{Comparison with the Lazarus-Leblond weight functions.}
\lb{compar}
\subsection{Local asymptotics.}
\lb{asymp}
Let us consider the Fourier transform of the
fundamental identity (\ref{ident})
\beq
\lb{ident2}
\overline{\sigma}^+_{i2} R_{ih} [\overline{U}^j_h]^+ - R_{ih}
\overline{\Sigma}^{j-}_{h2} [\overline{u}_i]^- = - \overline{p}_i
R_{ih} [\overline{U}^j_h]^+.
\eeq

Using the asymptotics for the physical fields and weight functions
given in the Appendix, we can observe that the structure of the
asymptotics for $\beta \in \Complex_{\beta}^+$ is as follows
\beq
\lb{leading}
\barr{l}
\overline{\sigma}^+ = A_{\sigma}\beta^{-1/2}
+ A^+_{\sigma}\beta^{-1/2+i\epsilon}
+ A^-_{\sigma}\beta^{-1/2-i\epsilon} + O(|\beta|^{-3/2}), \\[5mm]
[\overline{U}]^+ = A_{[U]}\beta^{-1/2} +
A^+_{[U]}\beta^{-1/2+i\epsilon} + A^-_{[U]}\beta^{-1/2-i\epsilon}
+ O(|\beta|^{-3/2}),
\earr
\quad \beta \to \infty,
\eeq
where
$A_{\sigma},A_{\sigma}^\pm$ and $A_{[U]},A_{[U]}^\pm$ are complex
quantities depending on $\lambda$.
Note that in equations  (\ref{leading}) and in equations
(\ref{leading2}) in the text below we suppress the subscript and
superscript indices (compare with the identity (\ref{ident2})) to
simplify notations. It follows that
\beq
\lb{as1}
\overline{\sigma}^+_{i2} R_{ih} [\overline{U}^j_h]^+
\sim \mA_{jk}(\lambda) \tilde{K}_k(\lambda) \frac{1}{\beta+i0},
\quad \beta \to \infty, \quad \beta \in \Complex_{\beta}^+,
\eeq
where
$
\{\tilde{K}_k(\lambda), k=1,2,3\} =
\{\tilde{K}(\lambda),\tilde{K}^*(\lambda),\tilde{K}_{\modIII}(\lambda)\}
$,
with the tilde denoting Fourier transform with respect to $x_3$
only,
$
{\mA}_{jk}(\lambda) =
\sqrt{|\lambda|} \{a_{jk}(\lambda)\} / (4\sqrt{2})
$
and
\beq
\barr{l}
\ds
a_{11}(\lambda) =
-\sign(\lambda)\frac{2d_*\sqrt{1-ia}}{c_1}
\frac{D}{|\lambda|^{i\epsilon}},
\qquad
\ds
a_{12}(\lambda) =
-\sign(\lambda)\frac{2d_*\sqrt{1-ia}}{c_2}
\frac{|\lambda|^{i\epsilon}}{D},
\\[5mm]
\ds
a_{21}(\lambda) =
\frac{d_*\sqrt{1-ia}}{c_1}\frac{D}{|\lambda|^{i\epsilon}},
\qquad
\ds
a_{22}(\lambda) =
-\frac{d_*\sqrt{1-ia}}{c_2}\frac{|\lambda|^{i\epsilon}}{D},
\\[5mm]
\ds
a_{31}(\lambda) =
\frac{2(1-ia)d_*^{3/2}}{(1+i)c_1 \sqrt{1-ad_*}}
\frac{D}{|\lambda|^{i\epsilon}},
\qquad
\ds
a_{32}(\lambda) =
\frac{2(1-ia)d_*^{1/2}\sqrt{1-ad_*}}{(1-i)c_2}
\frac{|\lambda|^{i\epsilon}}{D},
\\[5mm]
\ds
a_{13}(\lambda) =
2\sqrt{2}(1+i),
\qquad
\ds
a_{23}(\lambda) =
-\sign(\lambda)\frac{\sqrt{2}(1+i)(1+e_*)}{(i-a)ad_*^2},
\qquad
\ds
a_{33}(\lambda) = 0.
\earr
\eeq

We also note that, for $\beta \in \Complex_{\beta}^-$,
\beq
\lb{leading2}
\barr{l}
\overline{\Sigma}^-
= A_{\Sigma}\beta^{1/2} + A^+_{\Sigma}\beta^{1/2+i\epsilon}
+ A^-_{\Sigma}\beta^{1/2-i\epsilon} + O(|\beta|^{-1/2}), \\[5mm]
[\overline{u}]^- =
A_{[u]}\beta^{-3/2} + A^+_{[u]}\beta^{-3/2+i\epsilon}
+ A^-_{[u]}\beta^{-3/2-i\epsilon} + O(|\beta|^{-5/2}),
\earr
\quad
\beta \to \infty,
\eeq
and hence
\beq
\lb{as2}
R_{ih} \overline{\Sigma}^{j-}_{h2}
[\overline{u}_i]^- \sim \mA_{jk}(\lambda) \tilde{K}_k(\lambda)
\frac{1}{\beta-i0}, \quad \beta \to \infty, \quad \beta \in
\Complex_{\beta}^-.
\eeq
Then similar to Willis and Movchan (1995) we can write
\beq
\lb{delta}
\overline{\sigma}^+_{i2} R_{ih} [\overline{U}^j_h]^+ -
R_{ih} \overline{\Sigma}^{j-}_{h2} [\overline{u}_i]^- \sim
\mA_{jk}(\lambda) \tilde{K}_k(\lambda) \left(\frac{1}{\beta+i0} -
\frac{1}{\beta-i0}\right), \quad \beta \to \infty,
\eeq
where the
term in brackets can be identified as the regularization of the
delta function, namely $-2\pi i \delta(\beta)$.
Equation (\ref{ident2}) implies
\beq
\lim_{x_1' \to 0} \mF^{-1}_{x_1'}[\overline{\sigma}_{i2}^+ R_{ih}
[\overline{U}_h^j]^+ - R_{ih} \overline{\Sigma}^{j-}_{h2} [\overline{u}_i]^-]
=
-i
\mA_{jk}(\lambda) \tilde{K}_k(\lambda) = - \lim_{x_1' \to 0}
\mF^{-1}_{x_1'}[\overline{p}_i R_{ih} [\overline{U}_h^j]^+],
\eeq
and hence the Fourier transforms of the
stress intensity factors become
\beq
\lb{sif}
\tilde{K}_k(\lambda)
= -i \lim_{x_1' \to 0} \mA_{kj}^{-1}(\lambda) \int_{-\infty}^{0}
\tilde{p}_i(x_1,\lambda) R_{ih}
[\tilde{U}^j_h]^+(x_1'-x_1,\lambda) dx_1,
\eeq
where
$
\mA^{-1}_{kj}(\lambda) = \tilde{\mB}_{kj}(\lambda) =
\sqrt{2} \{\tilde{b}_{kj}(\lambda)\} / \left[
\sqrt{1-ia}(-ad_*+1+e_*)\right]
$
and
$$
\barr{l}
\ds
\tilde{b}_{11}(\lambda) =
-\sign(\lambda)\frac{c_1[-1+(-i+a)d_*](1+e_*)}{(-i+a)d_*^2}
\frac{|\lambda|^{i\epsilon}}{D\sqrt{|\lambda|}},
\qquad
\ds
\tilde{b}_{12}(\lambda) =
2c_1[-1+(-i+a)d_*]a
\frac{|\lambda|^{i\epsilon}}{D\sqrt{|\lambda|}},
\\[5mm]
\ds
\tilde{b}_{21}(\lambda) =
-\sign(\lambda)\frac{c_2[1+(-i+a)d_*](1+e_*)}{(-i+a)d_*^2}
\frac{D}{|\lambda|^{i\epsilon}\sqrt{|\lambda|}},
\qquad
\ds
\tilde{b}_{22}(\lambda) =
2c_2[1+(-i+a)d_*]a
\frac{D}{|\lambda|^{i\epsilon}\sqrt{|\lambda|}},
\earr
$$
\beq
\barr{l}
\ds
\tilde{b}_{31}(\lambda) =
-\frac{(1-i)}{\sqrt{2}}ad_*\frac{\sqrt{1-ia}}{\sqrt{|\lambda|}},
\qquad
\ds
\tilde{b}_{32}(\lambda) =
-\sqrt{2}\sign(\lambda)(1+i)(1+ia)ad_*^2\frac{\sqrt{1-ia}}{\sqrt{|\lambda|}},
\\[5mm]
\ds
\tilde{b}_{13}(\lambda) =
2ic_1[(1+ia)d_*^2+e_*]
\frac{|\lambda|^{i\epsilon}}{D\sqrt{|\lambda|}},
\qquad
\ds
\tilde{b}_{23}(\lambda) =
2ic_2[(1+ia)d_*^2-2-e_*]
\frac{D}{|\lambda|^{i\epsilon}\sqrt{|\lambda|}},
\\[5mm]
\ds
\tilde{b}_{33}(\lambda) =
-\sqrt{2}\sign(\lambda)(1+i)(-i+a)ad_*^3\frac{\sqrt{1-ia}}{\sqrt{|\lambda|}}.
\earr
\eeq
Note that the
components of the matrix $\mB$ are given by
$
\mB_{kj}(x_3) =
\sqrt{2}\{b_{kj}(x_3)\} /
\left[\sqrt{1-ia}(-ad_*+1+e_*)\right],
$
where
\beq
\barr{l}
\ds b_{11}(x_3) = - \frac{i}{\pi}\sign(x_3)
\frac{c_1[-1+(-i+a)d_*](1+e_*)}{(-i+a)d_*^2}
\frac{\cos[\pi(3+2i\epsilon)/4]\Gamma(1/2+i\epsilon)}
{D|x_3|^{i\epsilon}\sqrt{|x_3|}},
\\[5mm]
\ds
b_{12}(x_3) =
\frac{2}{\pi}c_1[-1+(-i+a)d_*]a
\frac{\cos[\pi(1+2i\epsilon)/4]\Gamma(1/2+i\epsilon)}
{D|x_3|^{i\epsilon}\sqrt{|x_3|}},
\\[5mm]
\ds
b_{21}(x_3) =
-\frac{i}{\pi}\sign(x_3)
\frac{c_2[1+(-i+a)d_*](1+e_*)}{(-i+a)d_*^2}
\cos[\pi(3-2i\epsilon)/4]\Gamma(1/2-i\epsilon)
\frac{D|x_3|^{i\epsilon}}{\sqrt{|x_3|}},
\\[5mm]
\ds
b_{22}(x_3) =
\frac{2}{\pi}c_2[1+(-i+a)d_*]a
\cos[\pi(1-2i\epsilon)/4]\Gamma(1/2-i\epsilon)
\frac{D|x_3|^{i\epsilon}}{\sqrt{|x_3|}},
\\[5mm]
\ds
b_{31}(x_3) =
-\frac{1-i}{2\sqrt{\pi}}ad_*\frac{\sqrt{1-ia}}{\sqrt{|x_3|}},
\qquad
\ds
b_{32}(x_3) =
\frac{i}{\sqrt{\pi}}
\sign(x_3)(1+i)(1+ia)ad_*^2\frac{\sqrt{1-ia}}{\sqrt{|x_3|}},
\\[5mm]
\ds
b_{13}(x_3) =
\frac{2i}{\pi}c_1[(1+ia)d_*^2+e_*]
\frac{\cos[\pi(1+2i\epsilon)/4]\Gamma(1/2+i\epsilon)}
{D|x_3|^{i\epsilon}\sqrt{|x_3|}},
\\[5mm]
\ds
b_{23}(x_3) =
\frac{2i}{\pi}c_2[(1+ia)d_*^2-2-e_*]
\cos[\pi(1-2i\epsilon)/4]\Gamma(1/2-i\epsilon)
\frac{D|x_3|^{i\epsilon}}{\sqrt{|x_3|}},
\\[5mm]
\ds
b_{33}(x_3) =
\frac{i}{\sqrt{\pi}}
\sign(x_3)(1+i)(-i+a)ad_*^3\frac{\sqrt{1-ia}}{\sqrt{|x_3|}}.
\earr
\eeq

\subsection{The integral representation of the Lazarus-Leblond weight functions.}
The  weight function $h_{kp}(x,z;x_3)$ $(k,p = 1,2,3)$,
corresponding to the work by Lazarus and Leblond (1998), are
defined as the k-th SIF generated at the point $x_3$ of the crack
front by a pair of point forces (of opposite direction) exerted on
the points $(x_1 = x, y = 0^{\pm}, x_3 = z)$ of the crack faces in
the directions $ \pm \be_p$.\footnote{Note that our notation is
not identical to the one  used
by Lazarus-Leblond. In fact $h_{1p}(x,z;x_3)$, $h_{2p}(x,z;x_3)$ are the
complex SIF and its conjugate, respectively, $h_{3p}(x,z;x_3)$ is the mode
III SIF. The Lazarus-Leblond's weight functions are given by
$$
h_{\modI p}(x,z;x_3) = \frac{h_{1p}(x,z;x_3) + h_{2p}(x,z;x_3)}{2},
~
h_{\modII p}(x,z;x_3) = \frac{h_{1p}(x,z;x_3) - h_{2p}(x,z;x_3)}{2i},
~
h_{\modIII p}(x,z;x_3) = h_{3p}(x,z;x_3).
$$}
Such a loading can be represented as
\beq
p_{ip}(x_1,x_3) =
-\delta_{ip} \delta(x_1-x) \delta(x_3-z),
\eeq
where $\delta_{ip}$
is the Kronecker delta. Taking the Fourier transform with respect
to $x_3$ we obtain
\beq
\tilde{p}_{ip}(x_1,\lambda) = -\delta_{ip}
\delta(x_1-x) e^{i\lambda z},
\eeq
and thus, it follows from
equation (\ref{sif}),
\beq
\tilde{h}_{kp}(x,z;\lambda) = i
e^{i\lambda z} \tilde{\mB}_{kj}(\lambda) \delta_{ip} R_{ih}
[\tilde{U}_h^j]^+(-x,\lambda).
\eeq
Finally, we deduce
\beq
\lb{integral}
h_{kp}(x,z;x_3) = \frac{1}{4\pi^2} \int_{-\infty}^{\infty}
\int_{-\infty}^{\infty} i \tilde{\mB}_{kj}(\lambda) \delta_{ip}
R_{ih} [\overline{U}_h^j]^+(\beta,\lambda)
e^{-i(-x)\beta}e^{-i(x_3-z)\lambda} d\lambda d\beta.
\eeq
The
functions $h_{kp}(x,z;x_3) \equiv h_{kp}(x,x_3-z)$ satisfy the parity and
homogeneity properties, as outlined by Lazarus and Leblond (1998):
\bit
\item parity
\bit
\item the functions $h_{1p},h_{2p}$ ($p=1,2$) and $h_{33}$ are even with respect
  to $x_3 - z$;
\item the functions $h_{13},h_{23}$ and $h_{3p}$ ($p=1,2$) are odd with respect
  to $x_3 - z$.
\eit
\item homogeneity
\bit
\item the functions $h_{1p}$ ($p=1,2,3$) are positively homogeneous of degree
  $-3/2 - i\epsilon$;
\item the functions $h_{2p}$ ($p=1,2,3$) are positively homogeneous of degree
  $-3/2 + i\epsilon$;
\item the functions $h_{3p}$ ($p=1,2,3$) are positively homogeneous of degree
  $-3/2$.
\eit
\eit

The constants $\gamma_+$, $\gamma_-$, $\gamma_{\modIII}$,
$\gamma_{z}$, $\gamma$ (see Section \ref{result}) are taken from
the asymptotics of the weight functions $h_{\modI j}, h_{\modII j}, h_{\modIII
j},$ as described in Section 3.5 of Lazarus and Leblond (1998).
Namely,
\beq
\barr{l}
\ds
[h_{\modI 2} + i h_{\modII 2}
+ i( h_{\modI 1} + i h_{\modII 1})](x, x_3-z)
\equiv \sqrt{\frac{|x|}{2 \pi}} |x|^{i \epsilon} H_+(x,x_3-z),
\\[5mm]
\ds
[h_{\modI 2} - i h_{\modII 2}
+ i( h_{\modI 1} - i h_{\modII 1})](x, x_3-z)
\equiv \sqrt{\frac{|x|}{2 \pi}} |x|^{i \epsilon} H_-(x,x_3-z),
\earr
\eeq
where
\beq
H_+(0, t) \equiv \gamma_+ |t|^{-2-2 i \epsilon}, \quad
H_-(0, t) \equiv \gamma_- t^{-2}.
\eeq
Similarly,
\beq
\barr{l}
\ds
[h_{\modIII 2} + i h_{\modIII 1}](x, x_3-z) \equiv
\sqrt{\frac{|x|}{2 \pi}} |x|^{i \epsilon} H_{\modIII}(x,x_3-z),
\\[5mm]
\ds
[h_{\modI 3} + i h_{\modII 3}](x, x_3-z) \equiv \sqrt{\frac{|x|}{2\pi}}
H_{z}(x, x_3-z),
\\[5mm]
\ds
h_{\modIII 3}(x, x_3-z) \equiv
\sqrt{\frac{|x|}{2 \pi}}  H(x,x_3-z),
\earr
\eeq
where
\beq
\lb{leb}
H_{\modIII}(0, t) \equiv \gamma_{\modIII} \sign(t)
|t|^{-2-2 i \epsilon}, \quad
H_{z}(0, t) \equiv
\gamma_{z} \sign(t) |t|^{-2-2 i \epsilon}, \quad
H(0,t) \equiv \gamma t^{-2}.
\eeq

We also note that one can write an equivalent representation
of $h_{kp}$ involving the matrix $\mB(x_3)$, defined in the text
of Section \ref{asymp}
\beq
h_{kp}(x,z;x_3) = i \{\mB_{kj}(\cdot) * \delta_{ip} R_{ih}
[U_h^j]^+(-x,\cdot)\}(x_3-z).
\eeq

This gives the exact integral representation of the
Lazarus-Leblond's weight functions, which further leads to the
formulae (\ref{gm1})--(\ref{gm5}). The procedure is
rather technical and it is outlined in Appendix (\ref{gamma}) for one of the
constants.

\subsection{Derivation of the constant $\gamma$.}
\lb{gamma}
In this section we will outline the general procedure to obtain the
Lazarus-Leblond constants from the formula (\ref{integral}). We will derive
in detail the formula for the constant $\gamma$ defined in (\ref{leb})$_3$; the
reasoning leading to the other formulae follows.

We note first that the homogeneity and parity properties of the function
$H(x,x_3-z)$ defined in (\ref{leb})$_3$ imply
\beq
\lb{eq45}
H(x,x_3) = \gamma f(-x/|x_3|) x_3^{-2},
\eeq
where $f(\zeta)$ is a smooth monotonic function on $\Reals_+$ with the
following properties:
\beq
\lb{prop}
f(0^+) = 1, \quad
f(\zeta) = f_0 \zeta^{-2} + O(\zeta^{-3}), \quad \text{as } \zeta \to +\infty,
\eeq
where $f_0$ is constant. When $\zeta < 0$, $f(\zeta)$ is defined as zero.

The property (\ref{prop})$_2$ follows immediately
from the fact that the stress intensity factors have to be finite as
$x_3 \to 0$ at any $x < 0$.

Note that $H(x,0) = \gamma f_0 / x^2$ and
\beq
h_{33}(x,0) = \sqrt{\frac{|x|}{2\pi}} H(x,0) =
\frac{\gamma}{\sqrt{2\pi}} f_0 |x|^{-3/2}.
\eeq

Before calculating the constant $\gamma$, we will derive the asymptotic
behaviour of the function $h_{33}(x,x_3) = \sqrt{|x|/(2\pi)} H(x,x_3)$ as
$x \to 0^-$. The function $h_{33}$ can be written in the equivalent form
\beq
h_{33}(x,x_3) = \frac{1}{\sqrt{|x|}} g(x,x_3), \quad
g(x,x_3) = \gamma \frac{|x|}{\sqrt{2\pi}} f(-x/|x_3|) x_3^{-2}.
\eeq
For any fixed $x_3 \ne 0$, we have
\beq
\lb{lim}
\lim_{x \to 0^-} g(x,x_3) = 0.
\eeq
On the other hand, for any $x < 0$,
\beq
\lb{int}
\int_{-\infty}^{\infty} g(x,x_3) dx_3 =
\frac{\gamma}{\sqrt{2\pi}}
\int_{-\infty}^{\infty} f(-x/|x_3|) \frac{|x|}{x_3^2} dx_3 =
\frac{2\gamma}{\sqrt{2\pi}} \int_{0}^{\infty} f(\zeta) d\zeta =
k_3^{(3)},
\eeq
where $k_3^{(3)}$ is a finite constant.
Then, (\ref{lim}) together with (\ref{int}) give
\beq
h_{33}(x,x_3) \sim \frac{k_3^{(3)}}{\sqrt{-x}} \delta(x_3), \quad x \to 0^-,
\eeq
which is consistent with the part (d) of the definition at the end of Section
\ref{fund_id}.

Using (\ref{integral}) we can write
\beq
\lb{h33}
h_{33}(x,x_3) =
\frac{1}{4\pi^2} \int_{-\infty}^{\infty}
\int_{-\infty}^{\infty} \frac{1}{\sqrt{|\lambda|}} F_{33}^+(\beta/|\lambda|)
e^{ix\beta}e^{-ix_3\lambda} d\lambda d\beta,
\eeq
where the function
\beq
F_{33}^+(\xi) = -i \sqrt{|\lambda|}
\{
\tilde{\mB}_{31}(\lambda) [\overline{U}^1_{*3}]^+(\xi,\lambda)
+ \tilde{\mB}_{32}(\lambda) [\overline{U}^2_{*3}]^+(\xi,\lambda)
+ \tilde{\mB}_{33}(\lambda) [\overline{U}^3_{*3}]^+(\xi,\lambda)
\}
\eeq
does not depend on $\lambda$, which is easy to check.

The function $F_{33}(\xi)$ is a ``+'' function and possesses the following
asymptotics
\beq
F_{33}^+(\xi) = (1+i) \xi^{-1/2} + \frac{(1-i)\pi}{4}\gamma_{33} \xi^{-3/2}
+ O(\xi^{-5/2}), \quad \xi \to \infty, \quad \xi \in \Complex^+_\xi,
\eeq
where
\beq
\gamma_{33} =
\frac{2}{\pi} \frac{3(1+e_*)-\sqrt{1-d_*^2}}{\sqrt{1-d_*^2}+1+e_*}.
\eeq

Now we have to investigate
\beq
\lb{gamma2}
\gamma = \lim_{x \to 0^-} H(x,x_3) x_3^2 =
\lim_{x \to 0^-} \sqrt{\frac{2\pi}{|x|}} h_{33}(x,x_3) x_3^2,
\eeq
for any fixed $x_3 \ne 0$.

Substituting $\xi = \beta/|\lambda|$ in (\ref{h33}), we obtain
\beq
\lb{h33b}
h_{33}(x,x_3) =
\frac{1}{4\pi^2} \int_{-\infty}^{\infty} \sqrt{|\lambda|} e^{-ix_3\lambda}
\int_{-\infty}^{\infty} F_{33}^+(\xi)
e^{ix \xi |\lambda|} d\xi d\lambda.
\eeq
Note that the integral in $\xi$ is the inverse Fourier transform of the
function $F_{33}^+(\xi) = \tilde{f}_{33}(\xi)$,
\beq
F_{33}^+(\xi) = \int_{-\infty}^{\infty} f_{33}(y) e^{i\xi y} dy,
\eeq
where
\beq
f_{33}(y) \equiv 0,  \quad \text{for } y < 0,
\eeq
and, by the Abelian theorem,
\beq
f_{33}(y) = \sqrt{\frac{2}{\pi}}\frac{1}{\sqrt{y}}
- \sqrt{\frac{\pi}{2}}\gamma_{33} \sqrt{y} + O(y^{3/2}), \quad y \to 0^+.
\eeq
Moreover, the function $f_{33}(y)$ decays at infinity sufficiently fast. Then
(\ref{h33b}) can be written in the form
\beq
h_{33}(x,x_3) = \frac{1}{2\pi}
\int_{-\infty}^{\infty} \sqrt{|\lambda|} e^{-ix_3\lambda} f_{33}(-x|\lambda|)
d\lambda.
\eeq
Note that $h_{33}(x,x_3) \equiv 0$ for $x > 0$, as expected. We also note that
the integrand is even with respect to $\lambda$ and hence
\beq
h_{33}(x,x_3) = \frac{1}{\pi}
\int_{0}^{\infty} \sqrt{\lambda} \cos(|x_3|\lambda) f_{33}(-x\lambda) d\lambda.
\eeq
Using the substitution $-x\lambda = y >0$ we deduce
\beq
\lb{eq81}
h_{33}(x,x_3) = \frac{1}{\pi (-x)^{3/2}}
\int_{0}^{\infty} \sqrt{y} \cos(y |x_3/x|) f_{33}(y) dy, \quad x < 0.
\eeq
Moreover, we can calculate the function $f(\zeta)$ defined in (\ref{eq45}) as
\beq
f(\zeta) = \left\{
\barr{ll}
\ds \sqrt{\frac{2}{\pi}} \frac{1}{\gamma \zeta^2}
\int_{0}^{\infty} \sqrt{y} \cos(y/\zeta) f_{33}(y) dy, & \zeta > 0, \\[5mm]
0, & \zeta \leq 0.
\earr
\right.
\eeq
Substituting (\ref{eq81}) into (\ref{gamma2}) we obtain
\beq
\gamma = \lim_{x \to 0^-} \sqrt{\frac{2}{\pi}} \left|\frac{x_3}{x}\right|^2
\int_{0}^{\infty} \sqrt{y} f_{33}(y) \cos(y|x_3/x|) dy,
\eeq
or, equivalently, for $x_3$ separated from zero, we can take the limit as
$t = |x_3/x| \to +\infty$
\beq
\gamma = \lim_{t \to \infty} \sqrt{\frac{2}{\pi}} t^{2}
\int_{0}^{\infty} \sqrt{y} f_{33}(y) \cos(ty) dy.
\eeq
Integrating by parts, we conclude
\beqar
\gamma
& = &
\lim_{t \to \infty} -\sqrt{\frac{2}{\pi}} t
\int_{0}^{\infty} [\sqrt{y} f_{33}(y)]' \sin(ty) dy \nonumber \\[5mm]
& = &
\lim_{t \to \infty} -\sqrt{\frac{2}{\pi}}
\left\{
[\sqrt{y} f_{33}(y)]'|_{y=0}
+ \int_{0}^{\infty} [\sqrt{y} f_{33}(y)]'' \cos(ty) dy
\right\} \\[5mm]
& = &
\lim_{y \to 0} -\sqrt{\frac{2}{\pi}} [\sqrt{y} f_{33}(y)]' = \gamma_{33},
\nonumber
\eeqar
which completes the required derivation.

\subsection{Comparison of the exact results and asymptotic approximations.}
The exact formulae (\ref{gm1})--(\ref{gm5}) are now compared with the
asymptotic approximations obtained in the earlier work by Lazarus and Leblond
(1998b) for small $\epsilon$:
\beq
\barr{l}
\ds
\gamma_+ = \frac{4\nu}{\pi(2-\nu)} + i\frac{8\nu \log 2}{\pi(2-\nu)}\epsilon
+ O(\epsilon^2),
\\[5mm]
\ds
\gamma_- = \frac{8(1-\nu)}{\pi(2-\nu)} - i\frac{16(1-\nu)}{\pi(2-\nu)}\epsilon
+ O(\epsilon^2),
\\[5mm]
\ds
\gamma_{\modIII} = -\frac{4(1-\nu)}{2-\nu}\epsilon
+ i\frac{4(1-\nu)(1-\log 2)}{2-\nu}\epsilon^2
+ O(\epsilon^3),
\\[5mm]
\ds
\gamma_z = \frac{4}{2-\nu}\epsilon + i\frac{4(1+\log 2)}{2-\nu}\epsilon^2
+ O(\epsilon^3),
\\[5mm]
\ds
\gamma = \frac{2(2+\nu)}{\pi(2-\nu)} + O(\epsilon^2),
\earr
\eeq
where
\beq
\nu = \frac{d^2+be}{b(b+e)}.
\eeq
It is convenient to use the notation
\beq
\eta = \frac{\mu_+ - \mu_-}{\mu_+ + \mu_-};
\eeq
we note that $\eta = 0$ when the shear moduli of the two elastic media become
equal.

\begin{figure}[!htb]
\begin{center}
\vspace*{3mm}
\includegraphics[width=15cm]{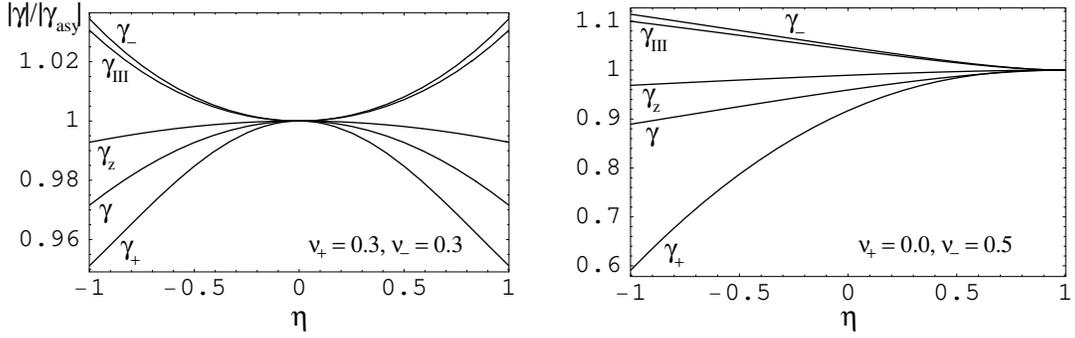}
\caption{\footnotesize Comparison of the moduli of the exact values of the
  Lazarus-Leblond constants to their asymptotic approximations.}
\label{fig03}
\end{center}
\end{figure}

For the sake of illustration, we present the results of comparison for the case
when $\nu_+ = \nu_- = 0.3$. We also consider an extreme situation when
$\nu_+ = 0$ and $\nu_- = 0.5$. Both diagrams are shown in Fig.\ \ref{fig03},
where we plot the modulus of the ratio of the exact values $\gamma_j$ to their
asymptotic approximations.

It is apparent that the asymptotic formulae by Lazarus and Leblond give a very
good agreement with the exact representations for the case of
$\nu_+ = \nu_- = 0.3$. It is also noted that similar accuracy is observed when
$0 \leq \nu_+ = \nu_- \leq 0.5$. The discrepancy becomes more pronounced when
$\nu_+$ and $\nu_-$ are not equal, which is illustrated in Fig.\ \ref{fig03}
for the case of $\nu_+ = 0$ and $\nu_- = 0.5$.

\begin{figure}[!htb]
\begin{center}
\vspace*{3mm}
\includegraphics[width=15cm]{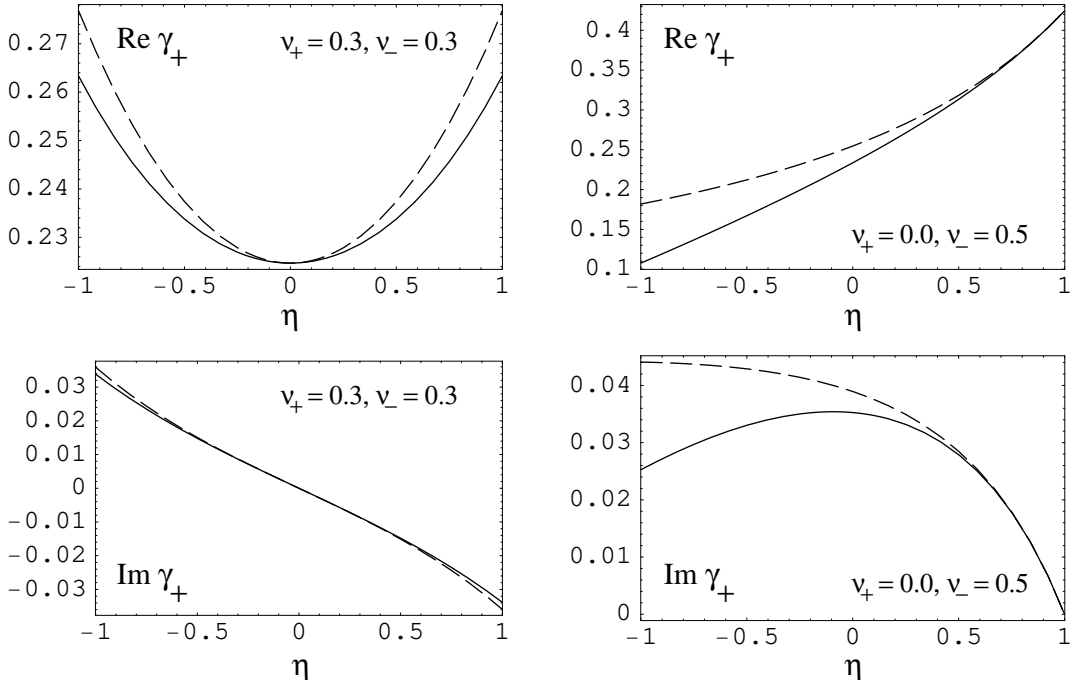}
\caption{\footnotesize Comparison of the exact results and asymptotic
  approximations for the constant $\gamma_+$.}
\label{fig04}
\end{center}
\end{figure}

The ``worst'' constant, in terms of its asymptotic approximation, is
$\gamma_+$.
Figure \ref{fig04} shows the results of computations for the same two cases as
in Fig.\ \ref{fig03} ($\nu_+ = \nu_- = 0.3$ and $\nu_+ = 0$, $\nu_- = 0.5$).
The exact values are shown by the solid line, whereas the dashed line shows
the corresponding asymptotic approximations.

\section{Discussion. Application to the wavy crack problem.}
\lb{discussion}
As follows from the earlier work by Willis and Movchan (1995),
Lazarus and Leblond (1998) and Bercial-Velez {\em et al.}\ (2005), the
asymptotic representation for the weight functions can be
efficiently used to evaluate the perturbation of the
stress intensity factors associated with a smooth perturbation of
the crack front (see the formulae in Section \ref{asymp} in the
text above). In particular, the identity (\ref{ident}), written in
terms of Fourier transforms, takes the form
\beq
\lb{BI}
\overline{\sigma}_{i2} R_{ih} [\overline{U}^j_h]^+ - R_{ih}
\overline{\Sigma}^{-j}_{h2} [\overline{u}_i] = - \overline{p}_i
R_{ih} [\overline{U}^j_h]^+,
\eeq
where the notations used for
components of tractions and displacements are the same as in the
text above.

Assume that the crack front is perturbed, within the plane $(x_1,
x_3),$ in such a way that \beq x_1 < \delta \phi(x_3), \quad x_2 =
0, \quad -\infty < x_3 < \infty, \eeq where $\delta$ is a small
positive parameter, and $\phi(x_3)$ is a smooth and bounded
function.

This induces  a small perturbation of tractions
${\Delta\sigma}_{i2}$ on the plane $x_2=0$ and associated
perturbation $[{\Delta u}_i]$ of the displacement jump across the
crack, whereas the components of tractions on the crack faces can
be assumed to remain unchanged. Then the corresponding identity
becomes
\beq
\lb{BI1}
(\overline{\sigma}_{i2} +
\overline{\Delta\sigma}_{i2}) R_{ih} [\overline{U}^j_h]^+ - R_{ih}
\overline{\Sigma}^{-j}_{h2} ([\overline{u}_i] + [\overline{\Delta
u}_i]) = - \overline{p}_i R_{ih} [\overline{U}^j_h]^+.
\eeq
Subtracting (\ref{BI}) from (\ref{BI1}) we obtain
\beq
\lb{ident3}
\overline{\Delta\sigma}_{i2} R_{ih} [\overline{U}^j_h]^+ - R_{ih}
\overline{\Sigma}^{-j}_{h2} [\overline{\Delta u}_i] = 0.
\eeq
Similar
to Section \ref{asymp}, one can analyse the delta function term in
the left-hand side of (\ref{ident3}) in order to obtain the formulae
for the stress intensity factors. As outlined in Willis and
Movchan (1995), Movchan {\em et al.} (1998) and Bercial-Velez {\em et
al.}\ (2005), the second-order terms in the asymptotic expansions
of components of the weight functions as well as components of
tractions near the crack front are essential in this asymptotic
analysis. The details of this asymptotic derivation are not the
purpose of the present paper though - the main result of the
present work is the derivation of the closed formulae for the
Lazarus-Leblond's constants (\ref{gm1})--(\ref{gm5}). Combined with
the asymptotic formulae of Section \ref{result}, they complete the
description of the stress intensity factors near the perturbed
edge of the interfacial crack.

\vspace{5mm}
{\footnotesize
{\em Acknowledgements--}
This research project has been supported by a Marie Curie Transfer of
Knowledge Grant of the European Communitys Sixth Framework Programme
under contract number (MTKD-CT-2004-509809).
The paper has been completed during the Marie Curie Fellowship of G. Mishuris
and A. Piccolroaz at Liverpool University. Provision of academic facilities by
the Department of Mathematical Sciences, Liverpool University, is gratefully
acknowledged.
The authors would like to thank Dr. J. Bercial-Velez for useful discussions on
the initial stage of this work.
}

\appendix
\renewcommand{\theequation}{\thesection.\arabic{equation}}

\section{APPENDIX.}
\setcounter{equation}{0}

\subsection{Solution of the Wiener-Hopf equation (\ref{whprob}).}
\underline{\em Factorization.} We use the factorization obtained
by Antipov (1999) for the matrix $\rho^{-1} \bG_0$,
\beq
\rho^{-1}
\bG_0(\beta,\lambda) = \bY_+^{-1}(\beta,\lambda)
\bY_-(\beta,\lambda),
\eeq
to obtain the factorization of the
matrix $\rho_*^{-1} \bG_*$
\beq
\rho_*^{-1}
\bG_*(\xi,\sign(\lambda)) = \bY_{*+}^{-1}(\xi,\sign(\lambda))
\bY_{*-}(\xi,\sign(\lambda)),
\eeq
where
\beq
\lb{whyp}
\bY_{*+} =
(\xi + i)^{1/2}
\left(
\barr{ccc}
0 & \ds
-\frac{\xi}{(1+e_*)\rho_*^2(\xi)}
& \ds \frac{\sign(\lambda)}{(1+e_*)\rho_*^2(\xi)} \\[5mm]
\ds \frac{\Sigma_{*1}^+(\xi)}{\delta_*(\xi)\Lambda_*^+(\xi)} & \ds
\frac{\sign(\lambda)\Sigma_{*2}^+(\xi)}{\delta_*(\xi)\Lambda_*^+(\xi)}
& \ds \frac{\xi \Sigma_{*2}^+(\xi)}{\delta_*(\xi)\Lambda_*^+(\xi)} \\[5mm]
\ds \frac{\rho_*^2(\xi)
\Sigma_{*2}^+(\xi)}{\delta_*(\xi)\Lambda_*^+(\xi)} & \ds
-\frac{\sign(\lambda)\Sigma_{*1}^+(\xi)}{\delta_*(\xi)\Lambda_*^+(\xi)}
& \ds -\frac{\xi \Sigma_{*1}^+(\xi)}{\delta_*(\xi)\Lambda_*^+(\xi)} \\[5mm]
\earr
\right),
\eeq
\beq
\lb{whym}
\bY_{*-} =
(\xi - i)^{-1/2}
\left(
\barr{ccc}
0 & \ds \frac{\xi}{\rho_*^2(\xi)}
& \ds -\frac{\sign(\lambda)}{\rho_*^2(\xi)} \\[5mm]
\ds -\frac{\sin(B_*^-(\xi))}{\rho_*(\xi)\Lambda_*^-(\xi)} & \ds
\frac{\sign(\lambda)\cos(B_*^-(\xi))}{\rho_*^2(\xi)\Lambda_*^-(\xi)}
& \ds \frac{\xi \cos(B_*^-(\xi))}{\rho_*^2(\xi)\Lambda_*^-(\xi)} \\[5mm]
\ds \frac{\cos(B_*^-(\xi))}{\Lambda_*^-(\xi)} & \ds
\frac{\sign(\lambda)\sin(B_*^-(\xi))}{\rho_*(\xi)\Lambda_*^-(\xi)}
& \ds \frac{\xi \sin(B_*^-(\xi))}{\rho_*(\xi)\Lambda_*^-(\xi)} \\[5mm]
\earr
\right),
\eeq
where
\beq
\lb{Bpm}
\barr{l}
\ds
\delta_*(\xi) = d_*^2\rho_*^2(\xi)-1,
\\[5mm]
\ds
B_*^{\pm}(\xi) = \rho_*(\xi)
\left[\frac{d_1}{2i} \psi_*^{\pm}(\xi) + B_{*1}^{\pm}(\xi)\right], \quad \xi
\in \Complex^\pm,
\\[5mm]
\ds
\psi_*^{\pm}(\xi) = \frac{i}{\pi \rho_*(\xi)}
\log\frac{\xi+\rho_*(\xi)}{\pm i}, \quad \xi
\in \Complex^\pm,
\\[5mm]
\ds
B_{*1}^{\pm}(\xi) =
-\frac{1}{4\pi} \int_{\Gamma}
\log\frac{d_*(t^2+1)^{1/2}+1}{d_*(t^2+1)^{1/2}-1}
\frac{dt}{(t^2+1)^{1/2}(t-\xi)}, \quad \xi
\in \Complex^\pm,
\earr
\eeq
and the contour of integration is as follow:
$$
\Gamma = \{ \xi = \xi_1 -i0, -\infty < \xi_1 \leq -|a| \} \cup
\{ \xi = \xi_1 +i0, |a| \leq \xi_1 < \infty \} \cup \{ \xi \in \Reals, |\xi|
< |a| \}.
$$
Other functions involved in the matrices (\ref{whyp})--(\ref{whym}) are defined
by
\beq
\barr{l}
\ds
\lb{sigmap}
\Sigma_{*1}^+(\xi) =
i d_* [\cos B_*^+ - i d_* \rho_* \sin B_*^+], \quad
\ds \Sigma_{*2}^+(\xi) =
i d_* [\frac{1}{\rho_*}\sin B_*^+ + i d_* \cos B_*^+],
\\[5mm]
\ds
\Lambda_*^+(\xi) =
\frac{d_0(\xi+i)^{1/2}}{(\xi+a)^{1/2}}, \quad \Lambda_*^-(\xi) =
\frac{(\xi-a)^{1/2}}{d_0(\xi-i)^{1/2}},
\earr
\eeq
where the constants used here and in the sequel are
\beq
\lb{const2}
d_0 = \sqrt[4]{1
- d_*^2}, \quad
d_1 = \log\frac{1 + d_*}{1 -
d_*}, \quad
a =
-\frac{\sqrt{1-d_*^2}}{d_*}, \quad
M_- = \frac{1}{d_0}
\sqrt{\frac{1-ia}{2}}.
\eeq
The branch cut for the logarithmic function in (\ref{Bpm})$_4$ is defined in the
same way as in Antipov (1999), formula (3.9).

The inverse of matrices (\ref{whyp}) and (\ref{whym}) are as
follow
\beq
\bY_{*+}^{-1} = (\xi + i)^{-1/2}
\left(
\barr{ccc}
0 &
\ds \frac{\Lambda_*^+(\xi)\Sigma_{*1}^+(\xi)}{d_*^2}
& \ds \frac{\Lambda_*^+(\xi)\Sigma_{*2}^+(\xi)}{d_*^2} \\[5mm]
-(1+e_*)\xi & \ds
\frac{\sign(\lambda)}{d_*^2}\Lambda_*^+\Sigma_{*2}^+ & \ds
-\frac{\sign(\lambda)}{d_*^2\rho_*^2}\Lambda_*^+\Sigma_{*1}^+ \\[5mm]
(1+e_*)\sign(\lambda) & \ds
\frac{\xi}{d_*^2}\Lambda_*^+\Sigma_{*2}^+ & \ds
-\frac{\xi}{d_*^2\rho_*^2}\Lambda_*^+\Sigma_{*1}^+ \\[5mm]
\earr
\right),
\eeq
\beq
\bY_{*-}^{-1} = (\xi - i)^{1/2}
\left(
\barr{ccc}
0 & \ds -\rho_*(\xi)\Lambda_*^-(\xi) \sin B_*^-(\xi)
& \ds \Lambda_*^-(\xi) \cos B_*^-(\xi) \\[5mm]
\ds \xi & \ds \sign(\lambda)\Lambda_*^-(\xi) \cos B_*^-(\xi) & \ds
\frac{\sign(\lambda)}{\rho_*(\xi)}\Lambda_*^-(\xi) \sin B_*^-(\xi) \\[5mm]
\ds -\sign(\lambda) & \ds \xi \Lambda_*^-(\xi) \cos B_*^-(\xi)
& \ds \frac{\xi}{\rho_*(\xi)} \Lambda_*^-(\xi) \sin B_*^-(\xi) \\[5mm]
\earr
\right).
\eeq

\underline{\em The solution.} Substituting the representation for
the matrix $\bG_*(\xi,\sign(\lambda))$ into the eq.\
(\ref{whprob}), we obtain
\beq
\bY_{*+}(\xi)
\mbox{\boldmath$\phi$}_{*}^+(\xi) = \bY_{*-}(\xi)
\mbox{\boldmath$\phi$}_{*}^-(\xi).
\eeq
Let us introduce the
notations
\beq
\bE_{*+} = \bY_{*+}(\xi)
\mbox{\boldmath$\phi$}_{*}^+(\xi), \quad \bE_{*-} = \bY_{*-}(\xi)
\mbox{\boldmath$\phi$}_{*}^-(\xi).
\eeq
It follows from the
asymptotics of the matrices $\bY_{*\pm}(\xi)$ at the points $\pm
a$ and $\pm i$ and from the asymptotics of the weight functions at
infinity that
\beq
\lb{entire}
\bE_{*+}(\xi) = \bE_{*-}(\xi) =
\left(
\barr{c}
\ds
\frac{C_1}{\xi-i} + \frac{C_2}{\xi+i} \\[5mm]
\ds
\frac{C_3}{\xi+i} + \frac{C_4}{\xi-a} \\[5mm]
\ds \frac{C_5}{\xi-a} + C_6
\earr
\right).
\eeq

Analyticity of the vectors $\mbox{\boldmath$\phi$}^{\pm}(\xi)$ at
the points $\pm i$ and $\pm a$ requires the following conditions
to be satisfied
\beq
\lb{cond}
\barr{l} \ds (1+e_*)\sign(\lambda)C_1 -
\frac{i}{2d_*M_-}
\left[\frac{C_5}{i-a} + C_6\right] = 0, \\[5mm]
C_2 + i \sign(\lambda) M_- C_3 = 0, \quad
i C_4 - d_* C_5 = 0.
\earr
\eeq
Note that the three linearly independent weight functions
(\ref{first})--(\ref{third}) have been obtained from (\ref{entire}), using
(\ref{cond}) and the following choice of the coefficients $C_j$:
\bit
\item[(1)] for the first weight function:

$C_1 = (1+e_*)^{-1}$,
$C_2 = C_3 = C_4 = C_5 = 0$, $C_6 = -2id_*M_-
\sign(\lambda)$,

\item[(2)] for the second weight function:

$C_1 = 0$, $C_2 = -id_*M_-^2 \sign(\lambda)$, $C_3 = d_*M_-$,
$C_4 = C_5 = C_6 = 0$,

\item[(3)] for the third weight function:

$C_1 = C_2 = C_3 = 0$, $C_4 =
-i M_-$, $C_5 = M_- d_*^{-1}$, $C_6 = -d_*^{-1} M_-
(i-a)^{-1}$.
\eit

\subsection{The asymptotics of weight functions.}
To obtain the
asymptotics of the weight functions, as $\xi \to \infty$, we will
make use of the following representations
\beq
\barr{l}
\ds
\psi_*^\pm(\xi) =
\frac{i}{\pi \xi} \log\frac{2\xi}{\pm i} + O\left(\frac{\log
\xi}{\xi^3}\right),
\\[5mm]
\ds
B_*^\pm(\xi) = \frac{d_1}{2\pi} \log \xi
+ \frac{d_1}{2\pi} \log \frac{2}{\pm i} + B_0
+ O\left(\frac{\log \xi}{\xi^2}\right),
\\[5mm]
\ds
\cos B_*^\pm(\xi) = c^\pm(\xi) + O\left(\frac{\log \xi}{\xi^2}\right), \quad
\sin B_*^\pm(\xi) = s^\pm(\xi) + O\left(\frac{\log \xi}{\xi^2}\right),
\earr
\qquad \xi \to \infty,
\eeq
where
\beq
\barr{l}
\ds
B_0 = \frac{1}{2\pi}
\int_0^\infty \log
\left|\frac{d_*(t^2+1)^{1/2}+1}{d_*(t^2+1)^{1/2}-1}\right|
\frac{dt}{(t^2+1)^{1/2}} - \frac{i}{2} \log \frac{1-ad_*}{d_*},
\\[5mm]
\ds
c^{\pm}(\xi) = \frac{1}{2} (e_0^{\pm 1} D \xi^{i\epsilon} +
e_0^{\mp 1} D^{-1} \xi^{-i\epsilon}), \quad s^{\pm}(\xi) =
-\frac{i}{2} (e_0^{\pm 1} D \xi^{i\epsilon} - e_0^{\mp 1} D^{-1}
\xi^{-i\epsilon}).
\earr
\eeq
We can show that
$$
\frac{1}{2\pi}
\int_0^\infty \log
\left|\frac{d_*(t^2+1)^{1/2}+1}{d_*(t^2+1)^{1/2}-1}\right|
\frac{dt}{(t^2+1)^{1/2}} = \frac{\pi}{4},
$$
so that
\beq
B_0 = \frac{\pi}{4} - \frac{i}{2} \log \frac{1-ad_*}{d_*}.
\eeq
In this case
\beq
\barr{l}
\ds
\xi F_1(\xi) \sim F_{10}(\xi) + \frac{1}{\xi} F_{11}(\xi), \quad
F_2(\xi) \sim F_{20}(\xi) + \frac{1}{\xi} F_{21}(\xi),
\\[5mm]
\ds
\xi P_1(\xi) \sim P_{10}(\xi) + \frac{1}{\xi}
P_{11}(\xi), \quad
P_2(\xi) \sim P_{20}(\xi) + \frac{1}{\xi} P_{21}(\xi),
\earr
\qquad \xi \to \infty,
\eeq
where
\beq
\barr{l}
\ds
F_{10}(\xi) = \sqrt{\frac{1-ia}{2}} (-id_* s^+(\xi)), \quad
F_{11}(\xi) = \sqrt{\frac{1-ia}{2}}
\left[c^+(\xi)-\frac{i}{2}d_*(i-a)s^+(\xi)\right],
\\[5mm]
\ds
F_{20}(\xi) = \sqrt{\frac{1-ia}{2}} (id_* c^+(\xi)), \quad
F_{21}(\xi) = \sqrt{\frac{1-ia}{2}}
\left[s^+(\xi)+\frac{i}{2}d_*(i-a)c^+(\xi)\right],
\\[5mm]
\ds
P_{10}(\xi) = \frac{1}{d_0^2}\sqrt{\frac{1-ia}{2}} s^-(\xi), \quad
P_{11}(\xi) = \frac{1}{d_0^2}\sqrt{\frac{1-ia}{2}}
\left(\frac{i-a}{2}\right) s^-(\xi),
\\[5mm]
\ds
P_{20}(\xi) = \frac{1}{d_0^2}\sqrt{\frac{1-ia}{2}} c^-(\xi), \quad
P_{21}(\xi) =
\frac{1}{d_0^2}\sqrt{\frac{1-ia}{2}} \left(\frac{i-a}{2}\right) c^-(\xi),
\earr
\eeq
and
\beq
e_0 = e^{\frac{\pi}{2}\epsilon}, \quad D
= 2^{i\epsilon}\frac{\sqrt{2}}{2} (1+i)
\sqrt{\frac{1-ad_*}{d_*}}.
\eeq
Finally, we can write the leading terms in asymptotics of the Fourier
transform of the weight functions,
which are of the order $O(\xi^{-1/2})$ and $O(\xi^{1/2})$, as $\xi \to \infty$,
for the displacement
and traction fields, respectively. Namely
\bit
\item[(1)] The first weight function:
\beq
\barr{l}
\ds
[\overline{U}_{*1}^1] \sim \sign(\lambda) \xi^{-1/2}
\left\{
-2 F_{10}(\xi) + \frac{1}{\xi}\left[1-2F_{11}(\xi)+iF_{10}(\xi)\right]
\right\}, \\[5mm]
\ds
[\overline{U}_{*2}^1] \sim \sign(\lambda) \xi^{-1/2}
\left\{
2 F_{20}(\xi) + \frac{1}{\xi}\left[2F_{21}(\xi)-iF_{20}(\xi)\right]
\right\}, \\[5mm]
\ds
[\overline{U}_{*3}^1] \sim -\xi^{-1/2}
\left\{
1+\frac{1}{\xi}\left[\frac{i}{2}+2F_{10}(\xi)\right]
\right\},
\earr
\eeq

\beq
\barr{l}
\ds
\overline{\Sigma}_{*12}^1 \sim -\sign(\lambda) \xi^{1/2}
\left\{
2id_* P_{10}(\xi)+\frac{1}{\xi}
\left[
\frac{1}{1+e_*}+2id_* P_{11}(\xi)+d_* P_{10}(\xi)
\right]
\right\}, \\[5mm]
\ds
\overline{\Sigma}_{*22}^1 \sim -\sign(\lambda) \xi^{1/2}
\left\{
2id_* P_{20}(\xi)+\frac{1}{\xi}
\left[
2id_* P_{21}(\xi)+d_* P_{20}(\xi)
\right]
\right\}, \\[5mm]
\ds
\overline{\Sigma}_{*32}^1 \sim \xi^{1/2}
\left\{
\frac{1}{1+e_*}+\frac{1}{\xi}
\left[
\frac{i}{2(1+e_*)}-2id_* P_{10}(\xi)
\right]
\right\}.
\earr
\eeq

\item[(2)] The second weight function:
\beq
\barr{l}
\ds
[\overline{U}_{*1}^2] \sim  \xi^{-1/2}
\left\{
iF_{20}(\xi)+\frac{1}{\xi}
\left[
\frac{1+e_*}{2d_0^2 d_* (i-a)}+iF_{21}(\xi)+\frac{3}{2}F_{20}(\xi)
\right]
\right\}, \\[5mm]
\ds
[\overline{U}_{*2}^2] \sim \xi^{-1/2}
\left\{
iF_{10}(\xi)+\frac{1}{\xi}
\left[
iF_{11}(\xi)+\frac{3}{2}F_{10}(\xi)
\right]
\right\}, \\[5mm]
\ds
[\overline{U}_{*3}^2] \sim \sign(\lambda) \xi^{-1/2}
\left\{
-\frac{1+e_*}{2d_0^2 d_* (i-a)}+\frac{1}{\xi}
\left[
iF_{20}(\xi)+\frac{3}{2}i\frac{1+e_*}{2d_0^2 d_* (i-a)}
\right]
\right\},
\earr
\eeq

\beq
\barr{l}
\ds
\overline{\Sigma}_{*12}^2 \sim \xi^{1/2}
\left\{
d_* P_{20}(\xi)+\frac{1}{\xi}
\left[
-\frac{1}{2d_0^2 d_* (i-a)} + d_* P_{21}(\xi)
-\frac{3}{2}id_* P_{20}(\xi)
\right]
\right\}, \\[5mm]
\ds
\overline{\Sigma}_{*22}^2 \sim \xi^{1/2}
\left\{
-d_* P_{10}(\xi)+\frac{1}{\xi}
\left[
-d_* P_{11}(\xi)
+\frac{3}{2}id_* P_{10}(\xi)
\right]
\right\}, \\[5mm]
\ds
\overline{\Sigma}_{*32}^2 \sim \sign(\lambda)\xi^{1/2}
\left\{
\frac{1}{2d_0^2 d_* (i-a)}+\frac{1}{\xi}
\left[
d_* P_{20}(\xi)-\frac{3i}{4d_0^2 d_* (i-a)}
\right]
\right\}. \\[5mm]
\earr
\eeq

\item[(3)] The third weight function:
\beq
\barr{l}
\ds
[\overline{U}_{*1}^3] \sim \frac{1}{d_*} \xi^{-1/2}
\left\{
F_{20}(\xi)+\frac{i}{d_* (i-a)} F_{10}(\xi) + \frac{1}{\xi}
\left[
F_{21}(\xi)+\frac{i}{d_* (i-a)}F_{11}(\xi)
\right.\right. \\[5mm]
\hspace{45mm}
\left.\left.
\ds
+\left(a-\frac{i}{2}\right)F_{20}(\xi)
+\left(\frac{3}{2}+ia\right)\frac{1}{d_* (i-a)}F_{10}(\xi)
\right]
\right\}, \\[5mm]
\ds
[\overline{U}_{*2}^3] \sim \frac{1}{d_*} \xi^{-1/2}
\left\{
F_{10}(\xi)-\frac{i}{d_* (i-a)} F_{20}(\xi) + \frac{1}{\xi}
\left[
F_{11}(\xi)-\frac{i}{d_* (i-a)}F_{21}(\xi)
\right.\right. \\[5mm]
\hspace{45mm}
\left.\left.
\ds
+\left(a-\frac{i}{2}\right)F_{10}(\xi)
-\left(\frac{3}{2}+ia\right)\frac{1}{d_* (i-a)}F_{20}(\xi)
\right]
\right\}, \\[5mm]
\ds
[\overline{U}_{*3}^3] \sim \frac{\sign(\lambda)}{d_*} \xi^{-1/2}
\left\{
\frac{1}{\xi}
\left[
F_{20}(\xi)+\frac{i}{d_* (i-a)}F_{10}(\xi)
\right]
\right\},
\earr
\eeq

\beq
\barr{l}
\ds
\overline{\Sigma}_{*12}^3 \sim \xi^{1/2}
\left\{
-iP_{20}(\xi)-\frac{P_{10}(\xi)}{d_* (i-a)}+\frac{1}{\xi}
\left[
-iP_{21}(\xi)-\left(\frac{1}{2}+ia\right)P_{20}(\xi)
\right.\right. \\[5mm]
\hspace{60mm}
\left.\left.
\ds
-\frac{1}{d_* (i-a)}P_{11}(\xi)-\frac{2a-3i}{2d_* (i-a)}P_{10}(\xi)
\right]
\right\}, \\[5mm]
\ds
\overline{\Sigma}_{*22}^3 \sim \xi^{1/2}
\left\{
iP_{10}(\xi)-\frac{P_{20}(\xi)}{d_* (i-a)}+\frac{1}{\xi}
\left[
iP_{11}(\xi)+\left(\frac{1}{2}+ia\right)P_{10}(\xi)
\right.\right. \\[5mm]
\hspace{60mm}
\left.\left.
\ds
-\frac{1}{d_* (i-a)}P_{21}(\xi)-\frac{2a-3i}{2d_* (i-a)}P_{20}(\xi)
\right]
\right\}, \\[5mm]
\ds
\overline{\Sigma}_{*32}^3 \sim \sign(\lambda) \xi^{1/2}
\left\{ \frac{1}{\xi}
\left[
-iP_{20}(\xi)-\frac{1}{d_* (i-a)}P_{10}(\xi)
\right]
\right\}.
\earr
\eeq
\eit

\subsection{Asymptotics of stress field.}
Stress components have the following asymptotics:
\beq
\barr{l}
\ds
\sigma_{12}
\sim -\frac{i}{2\sqrt{2\pi}} \left\{ K(x_3) x_1^{-1/2+i\epsilon} -
K^*(x_3) x_1^{-1/2-i\epsilon} +A(x_3) x_1^{1/2+i\epsilon} -
A^*(x_3) x_1^{1/2-i\epsilon} \right\} ,
\\[5mm]
\ds
\sigma_{22} \sim \frac{1}{2\sqrt{2\pi}} \left\{ K(x_3)
x_1^{-1/2+i\epsilon} + K^*(x_3) x_1^{-1/2-i\epsilon} +A(x_3)
x_1^{1/2+i\epsilon} + A^*(x_3) x_1^{1/2-i\epsilon} \right\},
\\[5mm]
\ds
\sigma_{32} \sim \frac{K_{\modIII}(x_3)}{\sqrt{2\pi}}
x_1^{-1/2} + \frac{A_{\modIII}(x_3)}{\sqrt{2\pi}} x_1^{1/2},
\earr
\eeq

The corresponding Fourier transforms are:
\beq
\barr{l}
\ds
\overline{\sigma}_{12} \sim \frac{1}{2} \left\{
\widetilde{K}(\lambda) \frac{1}{2 c_1 e_0} \beta^{-1/2 -
i\epsilon} - \widetilde{K}^*(\lambda) \frac{e_0}{2 c_2}
\beta^{-1/2 + i\epsilon} + \widetilde{A}(\lambda) \frac{1}{2 g_1
e_0} \beta^{-3/2 - i\epsilon} - \widetilde{A}^*(\lambda)
\frac{e_0}{2 g_2} \beta^{-3/2 + i\epsilon} \right\} ,
\\[5mm]
\ds
\overline{\sigma}_{22} \sim \frac{i}{2} \left\{
\widetilde{K}(\lambda) \frac{1}{2 c_1 e_0} \beta^{-1/2 -
i\epsilon} + \widetilde{K}^*(\lambda) \frac{e_0}{2 c_2}
\beta^{-1/2 + i\epsilon} + \widetilde{A}(\lambda) \frac{1}{2 g_1
e_0} \beta^{-3/2 - i\epsilon} + \widetilde{A}^*(\lambda)
\frac{e_0}{2 g_2} \beta^{-3/2 + i\epsilon} \right\},
\\[5mm]
\ds
\overline{\sigma}_{32} \sim \frac{1+i}{2}
\widetilde{K}_{\modIII}(\lambda) \beta^{-1/2} + \frac{-1+i}{4}
\widetilde{A}_{\modIII}(\lambda) \beta^{-3/2},
\earr
\eeq
where
\beq
\barr{l}
\ds
c_1
= \frac{(1+i)\sqrt{\pi}}{2\Gamma(1/2+i\epsilon)}, \quad c_2 =
\frac{(1+i)\sqrt{\pi}}{2\Gamma(1/2-i\epsilon)}, \quad e_0 =
e^{\epsilon \pi/2 },
\\[5mm]
\ds
g_1 =
\frac{(1-i)\sqrt{\pi}}{2\Gamma(3/2+i\epsilon)}, \quad g_2 =
\frac{(1-i)\sqrt{\pi}}{2\Gamma(3/2-i\epsilon)}.
\earr
\eeq

\subsection{Asymptotics of displacement components.}
The crack opening displacement are characterised by the following
asymptotics
\beqar
\lefteqn{ [u_1] \sim
-\frac{ib}{\sqrt{2\pi}\cosh({\pi\epsilon})} \left\{
\frac{K(x_3)}{1+2i\epsilon}(-x_1)^{1/2+i\epsilon}
-\frac{K^*(x_3)}{1-2i\epsilon}(-x_1)^{1/2-i\epsilon}
\right. } \nonumber \\
&& \hspace{65mm}
\left.\vphantom{\frac{K}{1}} +
B(x_3)(-x_1)^{3/2+i\epsilon}-B^*(x_3)(-x_1)^{3/2-i\epsilon}
\right\}, \nonumber
\eeqar

\beqar
\lefteqn{
[u_2] \sim
\frac{b}{\sqrt{2\pi}\cosh({\pi\epsilon})}
\left\{
\frac{K(x_3)}{1+2i\epsilon}(-x_1)^{1/2+i\epsilon}
+\frac{K^*(x_3)}{1-2i\epsilon}(-x_1)^{1/2-i\epsilon}
\right. } \\
&& \hspace{65mm}
\left.\vphantom{\frac{K}{1}}
+B(x_3)(-x_1)^{3/2+i\epsilon}+B^*(x_3)(-x_1)^{3/2-i\epsilon}
\right\}, \nonumber
\eeqar
\beq
[u_3] \sim \frac{2(b+e)}{\sqrt{2\pi}}
\left\{ K_{\modIII}(x_3)(-x_1)^{1/2}+B_{\modIII}(x_3)(-x_1)^{3/2}
\right\}. \nonumber
\eeq

The corresponding Fourier transforms are:
\beq
\barr{l}
\ds
[\overline{u}_1]
\sim -i \left\{ \widetilde{K}(\lambda)v_1e_0\beta^{-3/2-i\epsilon}
-\widetilde{K}^*(\lambda)\frac{v_2}{e_0}\beta^{-3/2+i\epsilon}
+\widetilde{B}(\lambda)w_1e_0\beta^{-5/2-i\epsilon}
-\widetilde{B}^*(\lambda)\frac{w_2}{e_0}\beta^{-5/2+i\epsilon}
\right\} ,
\\[5mm]
\ds
[\overline{u}_2] \sim
\widetilde{K}(\lambda)v_1e_0\beta^{-3/2-i\epsilon}
+\widetilde{K}^*(\lambda)\frac{v_2}{e_0}\beta^{-3/2+i\epsilon}
+\widetilde{B}(\lambda)w_1e_0\beta^{-5/2-i\epsilon}
+\widetilde{B}^*(\lambda)\frac{w_2}{e_0}\beta^{-5/2+i\epsilon},
\\[5mm]
\ds
[\overline{u}_3] \sim 2(b+e) \left\{
-\frac{1+i}{4}\widetilde{K}_{\modIII}(\lambda)\beta^{-3/2}
+\frac{3(-1+i)}{8}\widetilde{B}_{\modIII}(\lambda)\beta^{-5/2}
\right\},
\earr
\eeq
where
\beq
\barr{l}
\ds
v_1 = -\frac{ibd_0^2}{4c_1}, \quad v_2 =
-\frac{ibd_0^2}{4c_2},
\\[5mm]
w_1 =
\frac{ibd_0^2}{4g_1}\Gamma(3+2i\epsilon), \quad w_2 =
\frac{ibd_0^2}{4g_2}\Gamma(3-2i\epsilon).
\earr
\eeq

\end{document}